\documentclass[a4paper,11pt]{article}
\pdfoutput=1 

\usepackage{jheparxiv} 

\usepackage[T1]{fontenc} 

\usepackage{tensor}
\usepackage{amsmath}
\usepackage{amssymb}
\usepackage{mathrsfs}
\usepackage{graphicx}
\usepackage{stmaryrd}
\usepackage{twistor}
\usepackage{commath}
\usepackage{appendix}
\usepackage{yfonts}
\usepackage{tikz}
\usepackage{tikz-cd}
\usepackage{tkz-euclide,subfigure}
\usepackage{capt-of}
\usepackage{xfp} 
\usepackage[outline]{contour} 
\usetikzlibrary{decorations.markings,decorations.pathmorphing}
\usetikzlibrary{angles,quotes} 
\usetikzlibrary{arrows.meta}
\usepackage{tikz-3dplot}
\usetikzlibrary{patterns}
\usetikzlibrary{math}
\contourlength{1.4pt}

\newcommand{\veps}{\varepsilon}

\newcommand{\msf}[1]{\mathsf{#1}}

\newcommand{\kB}{\mathcal{B}}
\newcommand{\kA}{\mathcal{A}}
\newcommand{\kF}{\mathcal{F}}
\newcommand{\G}{\mathrm{G}}
 
\newcommand{\bz}{\bar z}
\newcommand{\dbeta}{\Dot{\beta}}
\newcommand{\myd}{\gamma}

\newcommand{\f}{\mathrm{f}}

\tikzset{>=latex} 
\colorlet{myred}{red!80!black}
\colorlet{myblue}{blue!80!black}
\colorlet{mygreen}{green!80!black}
\colorlet{mydarkred}{red!50!black}
\colorlet{mydarkblue}{blue!50!black}
\colorlet{mylightblue}{mydarkblue!6}
\colorlet{mypurple}{blue!40!red!80!black}
\colorlet{mydarkpurple}{blue!40!red!50!black}
\colorlet{mylightpurple}{mydarkpurple!80!red!6}
\colorlet{myorange}{orange!40!yellow!95!black}

\tikzstyle{world line}=[myblue!60,line width=0.4]
\tikzstyle{world line t}=[mypurple!60,line width=0.4]

\tikzset{declare function={%
  penrose(\x,\c)  = {\fpeval{2/pi*atan( (sqrt((1+tan(\x)^2)^2+4*\c*\c*tan(\x)^2)-1-tan(\x)^2) /(2*\c*tan(\x)^2) )}};%
  penroseu(\x,\t) = {\fpeval{atan(\x+\t)/pi+atan(\x-\t)/pi}};%
  penrosev(\x,\t) = {\fpeval{atan(\x+\t)/pi-atan(\x-\t)/pi}};%
  kruskal(\x,\c)  = {\fpeval{asin( \c*sin(2*\x) )*2/pi}};
}}

\def\Nsamples{40} 

\title{A hidden 2d CFT for self-dual Yang-Mills on the celestial sphere}

\abstract{Self-dual Yang-Mills theory admits an underlying infinite dimensional symmetry algebra, which has been obtained from mode expansion of Mellin transformed 4d scattering amplitudes and separately, Koszul duality on twistor space. In this paper, we propose to derive an explicit 2d realization of the algebra by performing a particular gauge transformation on the twistor action for self-dual Yang-Mills. The gauge parameter used in the transformation generates pure gauge connections corresponding to large gauge transformations on 4d Minkowski space, which localises part of the twistor action to a $\mathbb{CP}^1$ on $\mathbb{C}^*$ reduction of twistor space. Under a projection, it can be mapped to the celestial sphere at the light-cone cut of the origin on $\scri$. Geometrically, this is the common boundary celestial sphere shared by Euclidean AdS$_3$ or Lorentzian dS$_3$ slices of Minkowski space. We comment on the geometric meaning of the derivation from the perspective of minitwistor spaces of the 3d slices embedded in 4d Minkowski space. Using the action functional of this 2d CFT, we compute its stress-energy tensor and central charge. By a further marginal deformation, we calculate correlation functions of current algebra generators purely from the 2d side which reproduce 4d MHV form factors.}

\author[a]{Wei Bu,}\emailAdd{w.bu@sms.ed.ac.uk} 
\author[b]{Sean Seet}\emailAdd{sxes2@cam.ac.uk} 

\affiliation[{a}]{School of Mathematics and Maxwell Institute for Mathematical Sciences\\
University of Edinburgh, EH9 3FD, UK}

\affiliation[b]{Department of Applied Maths \& Theoretical Physics,\\
Wilberforce Road, Cambridge CB3 0WA,\\
United Kingdom}

\begin{document} 
\maketitle
\flushbottom

\section{Introduction}
The study of massless scattering amplitudes in 4d Minkowski space $\mathbb{M}$ using eigenstates which simultaneously diagonalise Lorentz boost and rotation has led to structures that resemble conformal correlation functions in a 2d conformal field theory (CFT) \cite{Pasterski:2016qvg,Pasterski:2017kqt}. This is especially evident when probing the infrared structures of 4d amplitudes, which exhibit the structure of operator product expansions (OPE) \cite{Fan:2019emx,Pate:2019lpp}. It is natural to ask if a 2d celestial conformal field theory (CCFT) does indeed exist on the celestial sphere at null infinity producing such dynamics. Even further, people have ventured in the literature to ask, given such correlations in observables and symmetries between 4d and 2d, to what extent one could interpret the putative CCFT as a codimension $2$ holographic dual for theories in 4d Minkowski space. 

An enormous amount of studies have been undertaken to probe the answer of this question. From the perspective of symmetry and kinematics, the exploration include studying symmetries of the Mellin transformed celestial amplitudes or constructing towers of conformally soft generators of infinite dimensional symmetry algebras \cite{Strominger:2021mtt,Guevara:2021abz,Fotopoulos:2020bqj,Fotopoulos:2019vac,Banerjee:2020kaa,Himwich:2021dau}. This approach heavily relied on the known scattering amplitudes in 4d Minkowski space, which made attempts to probe properties of the CCFT. Another perspective is provided by twistor theory, where the emergence of infinite-dimensional chiral algebras is explicitly linked to the classical integrability of the self-dual sector \cite{Adamo:2021lrv,Mason:2022hly,Monteiro:2022lwm}. Combining twistor theory with string theory also enables dynamical calculations of the the chiral algebras using the 2d worldsheet CFT \cite{Adamo:2021zpw,Adamo:2022wjo,Bu:2021avc,Adamo:2023zeh}, where in the MHV sector, the worldsheet is holomorphically identified with the celestial sphere. Although these are worldsheet CFTs which happen to be mapped to celestial sphere, one is reluctant to say that worldsheet theories are holographic (given this only works in a 4d/2d correspondence). Nevertheless, they provide suggestive hints for the form of the CCFT. More recently, an actual example of twisted holography in asymptotically locally Euclidean spacetimes was constructed \cite{Costello:2022jpg,Costello:2023hmi}. Although sufficiently close to flat space, the crucial difference is the presence of dimensionless scale parameter (number of branes), which determined the nature of the construction to be strong-weak. In genuine flat space, the absence of such parameters presents a challenge for such top-down constructions. \cite{Costello:2022wso} used Koszul duality to derive chiral OPEs underlying the S-algebra symmetry of SDYM, by requiring gauge invariant coupling of some abstract 2d chiral algebra and self-dual Yang-Mills (SDYM) in Euclidean twistor space of flat Minkowski space. No explicit construction of the chiral algebra was needed, given that they satisfy the S-algebra. Our goal here is to attempt to use the holographic reduction ideas \cite{deBoer:2003vf} to make case for a derivation of an explicit 2d CFT, whose correlation functions produce 4d dynamical observables. 

By reducing to 3d Euclidean AdS$_3$ (EAdS$_3$) or Lorentzian dS$_3$ (LdS$_3$) slices of 4d Minkowski space, with the well-established AdS$_3$/CFT$_2$ dictionary, one could hopefully reconstruct the original 4d theory using the 2d CFT \cite{Cheung:2016iub}. This was manifested in various ways, by considering coupling YM to a particular background scalar profile to regularise the scaling reduction \cite{Casali:2022fro}, from the embedding space formalism perspective \cite{Sleight:2023ojm,Iacobacci:2022yjo} and the space of oriented geodesics or (mini-)twistor spaces \cite{Bu:2023cef}. Based on the geometric framework developed in \cite{Bu:2023cef}, we would like to start with the twistor action of SDYM and derive an explicit 2d CFT after scaling reductions. Minitwistor space $\mathbb{MT}\cong\mathbb{CP}^1\times\mathbb{CP}^1\setminus\mathbb{CP}^1$ is the scaling reduction of twistor space $\mathbb{PT}^{\cO}$, being the space of oriented geodesics on EAdS$_3$, one lands on the boundary $S^2$ by degenerating the geodesics, which amounts going to the diagonal $\mathbb{CP}^1$ obtained by antiholomorphically identifying the two $\mathbb{CP}^1$ factors. This can be mapped to the celestial sphere at $u=0$ on $\scri^+$ or $v=0$ on $\scri^-$, where the light cone of the origin intersects $\scri$. In order to have a meaningful quantum field theory localising to this sphere, one needs fields which behave regularly in a large $r$ expansion, which are the asymptotically flat gauge fields. One particular example of such fields is a large gauge parameter\footnote{This is analogous (in spirit) to the edge mode construction on the level of on-shell soft wavefunctions in \cite{Kapec:2021eug}, although the separation of the soft sector was implemented by a cutoff in energy scale there.}. 

A large gauge transformation in 4d Minkowski space refers to the class of gauge transformations with $\cO(r^0)$ gauge parameters at null infinity. They are not quotiented out by the gauge fixing, but are physically significant field configurations that we will want to sum over. We take a particular class of large gauge parameters containing the $\Delta=1$ Goldstone as a special case. Their twistor space counterparts automatically localise the rest of the action functional to a $\mathbb{CP}^1$ after gauge transformations. Starting with SDYM on twistor space, we perform such gauge transformations and subsequently a $\mathbb{C}^*$ scaling reduction while keeping all the modes, which gives an minitwistor action corresponding to monopoles in EAdS$_3$ and a 2d CFT living on the celestial sphere. After recognizing the gauge parameters as quantum fields, we virtually obtain an equivalent description of SDYM on $\mathbb{MT}$ and the celestial sphere. Isolating the \emph{derived} action functional of the 2d CFT:
\begin{equation}\label{intro_cft}
    S_{\text{current}}=\int_{\mathbb{Z}+\im\mathbb{R}}\d^2\Delta\int_{\mathbb{CP}^1_{[\mu\bar\lambda]=0}}\tr\left(-\eta_{\Delta}\bar\partial\tilde\eta_{-\Delta}+a'_{\Delta} J_{-\Delta}+\tilde\phi_{\Delta}\bar\partial\phi_{-\Delta}+b'_{\Delta}\tilde J_{-\Delta}\right)\,.
\end{equation}
$\tilde\eta$ and $\tilde\phi$ are the large gauge parameters, $\phi$ and $\eta$ label the scalars obtained from scaling reduction while $a'$, $b'$ are the boundary modes of the positive/negative helicity gluons. They are all labeled by mode number $\Delta\in\mathbb{Z}+\im\mathbb{R}$. The currents are defined as integrals over the non-discrete modes:
\begin{align}
    &J^{\msf{a}}_{\Delta}=\int_{\mathbb{Z}+\im\mathbb{R}}\d^2j \,f^{\msf{a}}{}_{\msf{mn}}\left(:\eta^{\msf{m}}_{j}\tilde\eta^{\msf{n}}_{\Delta-j}: +:\phi^{\msf{m}}_{j}\tilde\phi^{\msf{n}}_{\Delta-j}:\right)-\Delta\tilde \phi_{\Delta}^{\msf{a}},\\
    &\tilde J^{\msf{a}}_{\Delta}=\int_{\mathbb{Z}+\im\mathbb{R}}\d^2j\,  f^{\msf{a}}{}_{\msf{mn}}\left(:\phi^{\msf{m}}_{j}\tilde\eta^{\msf{n}}_{\Delta-j}:\right)+(\Delta-2\im) \tilde \eta^{\msf{a}}_{\Delta}\,.
\end{align}
This set of generators give an explicit realisation of the S-algebra. Given the action functional for an irrational CFT (of infinite size) \eqref{intro_cft}, we also write down its stress-energy tensor, which agrees with the computations done in the literature \cite{Fan:2020xjj}, where the 2d stress tensor was constructed from double soft limit of bulk gluons. Its action on correlation functions was made manifest by taking Mellin transform of 4d scattering amplitudes in double soft limits. We also computed its central charge to be $\infty$, this agrees with the intuition that there are no dimensionless parameters in flat Minkowski space. 

Readers interested in the content of this 2d CFT are directed to section \ref{sec:CCFT}, its derivation relies on heavy twistor theory machinery laid out in section \ref{sec:large_gauge_4d} and \ref{sec:3d}, which can be skipped at first read. 

This paper is organized in the following way. In section \ref{sec:SDYM}, we review SDYM theory on spacetime and twistor space, after incorporating a term cancelling the 1-loop gauge anomaly on twistor space, we make statements which are quantum mechanically consistent. In section \ref{sec:large_gauge_4d}, we introduce the notion of large gauge transformations on spacetime, their uplifts to twistor space contain a free function which we later promote to a quantum field. After twistor space gauge transformations and appropriate field redefinitions, the SDYM action splits into two parts. In section \ref{sec:3d}, we scale reduce the split action and keep all the modes. One part of the action reduces to a minitwistor action for 3d monopoles while the other localises to the antiholomorphic diagonal $\mathbb{CP}^1$. In section \ref{sec:CCFT}, we focus on studying the 2d CFT action functional \eqref{intro_cft} by explicitly computing OPEs between the generators and the stress-energy tensor. In section \ref{sec_ff_cf}, in order to compute meaningful correlation functions using \eqref{intro_cft}, we marginally deform the CFT by inserting a local operator, which allows us to compute Yang-Mills MHV form factors purely from the 2d side. We comment on generalisations to N$^k$MHV and incorporation of the momentum conserving delta function by localising to formal neighborhoods of the $\mathbb{CP}^1$. We further discuss motivations of the form of the marginal operator in appendix \ref{sec:appendix_A} and explicit detail of the scaling reduction including $\mathbb{Z}+\im\mathbb{R}$ mode decompositions in appendix \ref{appendix:detail_scaling_reduction}.

\section{Quick review of self-dual Yang-Mills}\label{sec:SDYM}
\subsection{Spacetime self-dual Yang-Mills}
The action of Yang Mills theory on 4d Minkowski space $\mathbb{M}$ can be written as 
\begin{equation}
    \int_{\mathbb{M}}\tr\left( F^{\mu\nu}\, F_{\mu\nu}\right)\,\d^4x\,,
\end{equation}
where $F_{\mu\nu}$ is the skew-symmetric curvature 2-form.
Using the Pauli matrices $\sigma^{\mu}_{\alpha\dal}$ and the Lie algebra isomorphism $\mathfrak{so}(4,\mathbb{C})\cong\mathfrak{sl}(2,\mathbb{C})\otimes\mathfrak{sl}(2,\mathbb{C})$, one can replace the $\mathfrak{so}(4,\mathbb{C})$ index $\mu$ with $\mathfrak{sl}_2$ spinor indices $\alpha=0,1$ and $\dal=\Dot{0},\Dot{1}$. The skew-symmetry between $\mu$ and $\nu$ in $F_{\mu\nu}$ allows us to decompose it into two parts
\begin{equation}
    F_{\alpha\dal\beta\dbeta} = \epsilon_{\alpha\beta}F_{\dal\dbeta}+ \epsilon_{\dal\dbeta}F_{\alpha\beta}\,,
\end{equation}
with $\epsilon_{\alpha\beta}$ and $\epsilon_{\dal\Dot{\beta}}$ the usual 2d Levi-Civita tensors and $F_{\alpha\beta}= F^{\dal}{}_{\alpha\dal\beta}$, $F_{\dal\dbeta}= F^{\alpha}{}_{\dal\alpha\dbeta}$ the anti-self-dual/self-dual (ASD/SD) part of the curvature symmetric in its indices. Self-dual Yang-Mills (SDYM) refers to the action when $F_{\alpha\beta}=0$. With the spinor decomposition, its action is of BF-type:
\begin{equation}\label{SDYM_4d}
    \int_{\mathbb{M}}\tr\left(B^{\alpha\beta}\, F_{\alpha\beta}\right)\,\d^4x\,,
\end{equation}
where $B_{\alpha\beta}$ is some symmetric 2-form acting as a Lagrange multiplier, whose equation of motion gives the SDYM equation $F_{\alpha\beta}=0$ \footnote{$B_{\alpha\beta}$ represents the anti-self-dual degree of freedom.}. Gauge symmetry of the action is the standard 
\begin{equation}
    A_\mu \to A_\mu + \d_\mu\zeta,\quad B_\mu \to B_\mu -[B_{\mu},\zeta]+\d_{\mu}\xi\,.
\end{equation}

\subsection{Twistor space self-dual Yang-Mills}
Projective twistor space $\mathbb{PT}$ is the set $\mathbb{CP}^3\setminus\mathbb{CP}^1$:
\begin{equation}
    \mathbb{PT}:=\left\{Z^I=(\mu^{\dal},\lambda_{\alpha})\in\mathbb{CP}^3|\lambda_{\alpha}\neq 0 \right\}\,,
\end{equation}
where $Z^I=(\mu^{\dal},\lambda_{\alpha})$ are the four homogeneous coordinates on $\mathbb{CP}^3$ with $I={0,1,\Dot{0},\Dot{1}}$. They are related to complexified 4d Minkowski space $\mathbb{M}_{\mathbb{C}}\cong\mathbb{C}^4$ coordinatized by $x^{\alpha\dal}$ through the incidence relation:
\begin{equation}\label{incidence_relation}
    \mu^{\dal}=x^{\dal\alpha}\lambda_{\alpha}\,,
\end{equation}
which defines a $\mathbb{CP}^1_{x}\subset\mathbb{PT}$ for each point $x\in\mathbb{M}_{\mathbb{C}}$. In the rest of the paper, we shall remove a further $\mathbb{CP}^1$ from $\mathbb{PT}$ at $\mu^{\dal}=0$, which corresponds to removing the twistor line corresponding to the origin $0\in\mathbb{M}_{\mathbb{C}}$. We define:
\begin{equation}
    \mathbb{PT}^{\cO}:=\mathbb{CP}^3\setminus\{\{\lambda_{\alpha}=0\} \cup \{\mu^{\dal}=0\}\}\,.
\end{equation}
This breaks translation invariance explicitly, which corresponds to making an explicit choice of a degenerate `origin twistor' $O_{IJ}=\begin{pmatrix}
\veps_{\dal\Dot{\beta}} & 0 \\
0 & 0
\end{pmatrix}$. The explicit choice of origin also fixes the light-cone at $0\in\mathbb{M}^{3,1}$, which we shall use later to divide Minkowski space into Milne and Rindler regions. Classical SDYM theory on \eqref{SDYM_4d} can be described on $\mathbb{PT}^{\cO}$ by the BF-type action \cite{WARD197781,ward_wells,Boels:2006ir,Mason:2005zm}: 
\begin{equation}\label{SDYM_6d}
    S_{\text{SDYM}}=\int_{\mathbb{PT}^{\cO}}\D^3Z\wedge\tr\left(\kB\wedge \kF(\kA)\right)\,,
\end{equation}
where $\D^3Z=\epsilon^{IJKL}Z_I\,\d Z_J\wedge\d Z_K\wedge\d Z_L\in\Omega^{3,0}(\mathbb{PT}^{\cO},\cO(4))$ is the top holomorphic form on $\mathbb{PT}^{\cO}$. $\kB\in \Omega^{0,1}(\mathbb{PT}^{\cO},\cO(-4)\otimes\mathfrak{g})$\footnote{The full non-linear construction of Ward bundle $E$ requires the fields to take values in endomorphisms of the bundle. One identifies End$(E)$ with $\mathfrak{g}$ after holomorphic trivialisation.} is a $(0,1)$-form valued in the Lie algebra of the gauge bundle, $\kF=\bar\partial \kA+\kA\wedge \kA$ with $\kA\in \Omega^{0,1}(\mathbb{PT}^{\cO},\cO\otimes\mathfrak{g})$ the holomorphic part of the connection on the gauge bundle over twistor space. With appropriate choice of gauge, \eqref{SDYM_6d} can be shown to reduce to SDYM on $\mathbb{M}$ \eqref{SDYM_4d}. 

The gauge transformations of the BF twistor action \eqref{SDYM_6d} take the following form: 
\begin{equation}\label{GT_6d}
    \kB \rightarrow \kB - [\kB,\Lambda] + \bar \D \chi, \quad \kA \rightarrow \kA + \bar \D \Lambda\,,
\end{equation}
where $\bar\D=\bar\partial+\kA$ is the covariant anti-holomorphic derivative. $\chi\in\Omega^0(\mathbb{PT}^{\cO},\cO(-4)\otimes\mathfrak{g})$ and $\Lambda\in\Omega^0(\mathbb{PT}^{\cO},\cO\otimes\mathfrak{g})$ are the gauge parameters with appropriate form degree and homogeneous scaling weights. Note that there is more gauge redundancy on $\mathbb{PT}^{\cO}$ than on spacetime, due to the presence of extra spinor fibres. 

It was first pointed out in \cite{costello2021quantizing} that SDYM action \eqref{SDYM_6d} suffers from a 1-loop gauge anomaly $\int_{\mathbb{PT}^{\cO}}\tr\left(\Lambda\,\left(\partial\kA\right)^3\right)$. With gauge groups obeying the quadratic/quartic trace factorisation identity $\lambda_{\mathfrak{g}}^2\mathrm{Tr} (X^2)^2 = \tr(X^4)$, this anomaly can be compensated by the addition of an $(1,1)$-form that will correspond to an axion on spacetime. We denote traces over the fundamental with $\mathrm{Tr}$ and those over the adjoint indices with $\tr$. Integrating out the axion gives the following counter term:
\begin{equation}\label{S_counter}
    S_{\text{ct}}[\kA] :=\int_{\mathbb{PT}^{\cO}} \mathrm{Tr}\left(\kA\partial \kA\right)\partial\frac{1}{\bar\partial}\mathrm{Tr}\left(\kA\partial \kA\right) \,,
\end{equation}
with appropriate normalization factor, the usual gauge transformation \eqref{GT_6d} of \eqref{S_counter} cancels the gauge anomaly\footnote{\eqref{S_counter} can be matched onto the non-local term in \cite{costello2021quantizing} by explicitly writing the axion propagator in affine coordinates, which amounts to a double pole.}.

In Woodhouse gauge \cite{Woodhouse:1985id}, $S_{\text{ct}}$ vanishes and we obtain SDYM on spacetime. If one were to recover the axion, SDYM acquires a quantum effective correction which cancels the one-loop all plus amplitudes. Since SDYM theory is 1-loop exact, we shall take the quantum mechanically consistent SDYM to be on twistor space
\begin{equation}\label{QSDYM_6d}
    S_{\text{QSDYM}} = \int_{\mathbb{PT}^{\cO}}\D^3Z \wedge\tr\left(\kB\wedge \kF(\kA)\right) -\mathfrak{C}_{\text{YM}}^2\mathrm{Tr}\left(\kA\partial \kA\right)\partial\frac{1}{\bar\partial}\mathrm{Tr}\left(\kA\partial \kA\right)+\cO(\kA^5)\,,
\end{equation}
where $\mathfrak{C}_{\text{YM}}$ is calculated in \cite{costello2021quantizing} to be:
\begin{equation}
    \mathfrak{C}_{\text{YM}}= \sqrt{\frac{1}{(2\pi)^4\, 32}}\,.
\end{equation}
And the the $\cO(\kA^5)$ terms formally denote higher multiplicity vertices built from Feynman rules of $S_{\text{SDYM}}$ and $S_{\text{ct}}$. In order to interpret the non-local term as an axion coupling, we must restrict ourselves to gauge groups satisfying the trace factorisation identity.
\section{Large gauge transformation in 4D}\label{sec:large_gauge_4d}
\subsection{Spacetime Large gauge transformation (4D picture)}
On 4d Minkowski space $\mathbb{M}$, one is allowed to take \textit{large gauge transformations} \cite{Strominger:2013lka,Kapec:2014zla}, where the gauge parameter behaves regularly asymptotically on $\scri^+$.
More concretely, we can coordinatize $\scri^+$ using homogeneous coordinates $(\lambda,\bar\lambda)$ on the celestial sphere and retarded time $u$. After conformal compactification, the degenerate metric at large $r$ becomes
\begin{equation}\label{scri_metric}
    \d s^2= 0\times\d u^2+\frac{\D\lambda\wedge\D\bar\lambda}{\la\lambda\hat{\lambda}\ra^2}\,,
\end{equation}
where $\hat{\lambda}_{\alpha}=\binom{-\bar\lambda_1}{\bar\lambda_0}$ and $\bar\lambda_{\dal}=\binom{\bar\lambda_0}{\bar\lambda_1}$ define conjugations of the two spinor $\lambda_{\alpha}=\binom{\lambda_0}{\lambda_1}$ and $0$ indicates the $\d u$ component is in fact subleading of order $\cO(1/r)$.

One could take the spacetime connection 1-form $A$ and decompose it under a large $r$ expansion, up to order $\cO(1/r)$ (under partial gauge fixing $A_r=0$ in radial gauge):
\begin{equation}
    \left.A\right\vert_{r\to\infty} = A_{\lambda}\D\lambda+A_{\bar\lambda}\D\bar\lambda+\mathcal{O}\left(\frac{1}{r}\right) \,,
\end{equation}
where we have in addition fixed temporal gauge where the $u$-component vanishes on $\scri^+$. Large gauge transformations in 4d then refer to gauge transformations which take the functional degrees of freedom $A_\lambda$ or $A_{\bar\lambda}$ as gauge parameters. We could give a more unified definition for these functional degrees of freedom by noting that asymptotic curvatures are essentially flat\footnote{In other words, the connection at $\scri^+$ is pure gauge.}. This allows us to write an analytic function $\G(\lambda,\bar\lambda)$ which does not depend on $u$ (allowing poles at isolated points) such that:
\begin{equation}
   A_{\lambda}\D\lambda\to A_{\lambda}\D\lambda+\G\D\lambda , \quad  A_{\bar\lambda}\D\bar\lambda\to A_{\bar\lambda}\D\bar\lambda+\G\D\bar\lambda \,,
\end{equation}
as a gauge transformation of the asymptotic components of the gauge connection.
The discussions above can be repeated for $\scri^-$ in similar manner, with retarded time $u$ replaced by advanced time $v$.

Hence taking a large gauge transformation amounts to taking gauge parameters to be such $\G$s, with gauge transformations $A_{\mu}\to A_{\mu}+\partial_\mu \G$. The essential requirement for such $\G$s is that they fall off like $\cO(r^0)$, or that they scale like $x^0$. Gauge transformations using a subset of these pure gauge large gauge parameters $\G$ have been identified as the $\Delta=1$ Goldstone modes in conformal primary basis \cite{Donnay:2018neh}, which take the form:
\begin{equation}\label{Delta=1_gold}
    A_\mu^{1} = \frac{\partial}{\partial x^\mu}\,\left(-\frac{x\cdot\veps}{x\cdot p}\right)\,,
\end{equation}
with $p^{\mu}$ the external momenta and $\veps^{\mu}$ the polarization vector satisfying $\veps\cdot p=0$.

\subsection{Large gauge transformations on twistor space}
Similarly, one could take a holomorphic gauge transformation \eqref{GT_6d} on twistor space using some `large gauge parameter'. Although the notion of large $r$ expansion is less intuitively clear on $\mathbb{PT}^{\cO}$\footnote{It is clear $r^2=2[\mu\hat{\mu}]/\la\lambda\hat{\lambda}\ra$ considering the Euclidean real slice of $\mathbb{PT}^{\cO}$, but in the complexified setting, the physical meaning of large $r$ expansion is obscured on twistor space.}, the constraint that we are only interested in the piece homogeneous of degree 0 in $x$ lets us constrain ourselves to consider gauge parameters homogeneous of total degree 0 in $\mu$ and $\bar \mu$. The subset of large gauge transformations that we will consider are those which are permitted to be singular only on $[\mu \bar \lambda]=0$ (see Appendix \ref{appendix lgt}). The simplest nontrivial example is the following $\bar \partial$-exact representative $\kA\in \Omega^{0,1}(\mathbb{PT}^{\cO},\cO\otimes\mathfrak{g})$ (we suppress all the adjoint indices as they will play no role):
\begin{equation}\label{PT_1_gold}
    \kA = \bar\partial\left(\frac{\la\lambda\hat{\lambda}\ra[\mu\hat{\bar\lambda}]}{[\mu\bar\lambda]}\,\tilde\eta\right)\,,
\end{equation}
where $\hat{\bar\lambda}_{\dal}= \binom{-\lambda_1}{\lambda_0}$ defines conjugation of the two-spinor $\lambda_{\alpha}=\binom{\lambda_0}{\lambda_1}$. $\tilde\eta$ is a section of $\mathcal{O}(-2)$ taking values in the lie algebra $\mathfrak{g}$. We ask $\tilde\eta$ to be of homogeneity $0$ in $\mu$ and behave regularly on $[\mu\bar\lambda]=0$, which deems its spacetime counterpart $\tilde\eta(\kappa,x)$ to have homogeneity $0$ in $x$. In other words, we only consider those $\tilde\eta(\kappa,x)$ which behave regularly in a large $r$ expansion and does not blow up at the locus $[\mu\bar\lambda]=0$. With non-trivial choices of $\tilde\eta$, \eqref{PT_1_gold} represents a generalisation of the $\Delta=1$ Goldstone on $\mathbb{PT}^{\cO}$. which can be verified through an integral transform to spacetime\footnote{This is the Sparling transform \cite{Sparling1990}, equivalent to the usual Penrose transform \cite{Bu:2023cef}, where different choices of $\iota_\alpha$ differ by $\bar\partial$-exact terms in the integrand..}:
\begin{equation}
    A_{\alpha\dal}(x)=\int_{\mathbb{CP}^1}\D\lambda\wedge\frac{\iota_\alpha}{\la\lambda\iota\ra}\,\left.\frac{\partial}{\partial \mu^{\dal}}\left(\bar \partial \left(\frac{\la\lambda\hat{\lambda}\ra[\mu\hat{\bar\lambda}]}{[\mu\bar\lambda]}\,\tilde\eta\right)\right)\right\vert_{\mu=x\lambda}\,,
\end{equation}
where $\iota_{\alpha}$ is a reference spinor and we have restricted the integrand at the projective line using the incidence relation \eqref{incidence_relation}. The integral can be evaluated by integrating by parts on $\bar\partial$, which only picks up the $\la\lambda\iota\ra$ pole in front and give the following holomorphic delta function:
\begin{equation}
    \bar\delta(\la\lambda\iota\ra)=\frac{1}{(2\pi\im)^2}\,\bar\partial\left(\frac{1}{\la\lambda\iota\ra}\right) \,.
\end{equation}
Implementing this, we have
\begin{equation}\label{eq3}
    A_{\alpha\dal}(x)= -\int_{\mathbb{CP}^1}\D\lambda\wedge\bar\delta(\la\lambda\,\iota\ra)\frac{\la\iota\,\hat \iota\ra}{\la\lambda\,\hat \iota\ra}\frac{\partial}{\partial x^{\alpha\dal}}\left(\frac{\la\lambda\hat{\lambda}\ra\,[\mu\hat{\bar\lambda}]}{[\mu\bar\lambda]}\,\tilde\eta(\mu^{\dal},\lambda_{\alpha})\right) 
    = - \frac{\partial}{\partial x^{\alpha\dal}}\,\left(\frac{x\cdot \veps}{x\cdot p} \,\tilde{\boldsymbol{\eta}}(x,\kappa) \right)\,,
\end{equation}
where $\hat{\iota}_{\alpha}=\binom{-\bar\iota_1}{\bar\iota_0}$ is the quaternionic conjugate of $\iota_{\alpha}=\binom{\iota_0}{\iota_1}$ and we used $\boldsymbol{\tilde\eta}$ to denote the spacetime counterpart of $\tilde\eta$ on $\mathbb{PT}^{\cO}$. We have picked $\iota_\alpha=\kappa_\alpha$ and $\bar\iota_{\dal}=\tilde\kappa_{\dal}$ in the last step, with $p_{\alpha\dal}=\kappa_{\alpha}\tilde\kappa_{\dal}$ the usual spinor helicity parametrization for external null momenta. $\veps_{\alpha\dal}= \kappa_{\alpha}\hat{\tilde\kappa}_{\dal}$ denotes polarization vector which contracts with $p_{\alpha\dal}$ to $0$. This exactly agrees with the pure gauge $\Delta=1$ Goldstone mode \eqref{Delta=1_gold} when we pick $\tilde{\boldsymbol{\eta}}(x,\kappa)=1$ on spacetime. 

Since \eqref{PT_1_gold} descends to a pure gauge large parameter on $\mathbb{M}$, we would like to naturally define a `large gauge transformation' on $\mathbb{PT}^{\cO}$ to be a gauge transformation \eqref{GT_6d} with gauge parameters $\Lambda=\frac{\la\lambda\hat{\lambda}\ra[\mu\hat{\bar\lambda}]}{[\mu\bar\lambda]}\,\tilde\eta$ with $\tilde \eta$ obeying the homogeneity condition but permitted to be singular. If we want to be fully explicit about the order of the pole at $[\mu \bar \lambda]$, we would decompose in a sum of terms like the following $\Omega^{0,1}(\mathbb{PT}^{\cO},\cO\otimes\mathfrak{g})$ with $\myd\in\mathbb{Z}_+$: 
\begin{equation}\label{eq2}
    \kA=\bar \partial \left(\left(\frac{\la\lambda\hat{\lambda}\ra\,[\mu\hat{\bar\lambda}]}{[\mu\bar\lambda]}\right)^{\myd} \,\tilde\eta\right)\,,
\end{equation}
where $\tilde\eta\in\Omega^0(\mathbb{PT}^{\cO},\cO(-2\myd)\otimes\mathfrak{g})$ and required to be smooth. They are uplifts of the following pure gauge connection 1-forms on $\mathbb{M}$: 
\begin{equation}\label{eq1}
   A_{\alpha\dal}(x) = - \frac{\partial}{\partial x^{\alpha\dal}}\,\left(\left(\frac{x\cdot \veps}{x\cdot p}\right)^{\myd} \,\boldsymbol{\tilde\eta}(\kappa,x) \right),\quad \myd\in\mathbb{Z}_+\,,
\end{equation}
which are also evidently pure gauge modes not vanishing at infinity given $\boldsymbol{\tilde\eta}$ is of homogeneity $0$ in $x$. Note that these have not been recognized in the literature as any particular spin-1 conformal primary in the putative celestial conformal field theory. \eqref{eq1} are pure gauge with homogeneity $0$ in $x$ in the gauge parameter part, hence they are all large gauge components behaving regularly at infinity. There is an enormous class of such connection 1-forms on $\mathbb{M}$, the $\Delta=1$ Goldstone and the representative of interest here are just simple examples.

Given these profiles on $\mathbb{PT}^{\cO}$ which correspond to spacetime to large gauge transformations, `large gauge transformations' on $\mathbb{PT}^{\cO}$ are generated by taking the gauge parameters $\Lambda=\left(\frac{\la\lambda\hat{\lambda}\ra\,[\mu\hat{\bar\lambda}]}{[\mu\bar\lambda]}\right)^{\myd} \,\tilde\eta$ in \eqref{GT_6d}:
\begin{equation}\label{PT_lgt_explicit}
    \kA \rightarrow \kA + \bar \D \left( \left(\frac{\la\lambda\hat{\lambda}\ra\,[\mu\hat{\bar\lambda}]}{[\mu\bar\lambda]}\right)^{\myd} \tilde \eta\right)\,.
\end{equation}
Evaluating the covariant derivatives of gauge parameters of arbitrary $\myd$:
\begin{multline}\label{barD_analysis}
    \bar\D\Lambda = \sum_{\myd}\bar\partial\left(\frac{1}{[\mu\bar\lambda]^{\myd}}\right)\tilde\eta_{\myd} ([\mu\hat{\bar\lambda}]\la\lambda\hat{\lambda}\ra)^{\myd}+\left(\frac{[\mu\hat{\bar\lambda}]\la\lambda\hat{\lambda}\ra}{[\mu\bar{\lambda}]}\right)^{\myd}\left(\bar\partial\tilde\eta_{\myd}-\myd\frac{[\mu\hat{\bar\lambda}]}{[\mu\bar{\lambda}]}\frac{\D\hat{\lambda}}{\la\lambda\hat{\lambda}\ra} \tilde\eta_{\myd}+[\kA,\tilde\eta_{\myd}]\right)\,,
\end{multline}
where we have isolated terms when $\bar\partial$ points along $\bar\mu$ in the first term and picks up the pole $1/[\mu\bar\lambda]^{\myd}$. Note that the first term resembles a holomorphic delta function localising the rest of the integrand to the locus of $\{[\mu\bar\lambda]=0\}$ while the other terms do not. We can absorb the non-localising terms as an additive field redefinition $\kA\to\kA'$. Geometrically, $\bar\partial(1/[\mu\bar\lambda]^{\myd})=\bar\delta^{(\myd-1)}([\mu\bar\lambda])$ localises the action to the $(\myd-1)$th formal neighborhood of $\mathbb{CP}^1_{[\mu\bar\lambda]}\subset\mathbb{PT}^{\cO}$. 

Including non-localising terms in \eqref{barD_analysis}, we do the following field redefinition $\kA\to\kA'$ with
\begin{equation}\label{field_redefinition}
    \kA'
    =\kA+  \sum_{\myd\in\mathbb{Z}_+}\left(\frac{[\mu\hat{\bar\lambda}]\la\lambda\hat{\lambda}\ra}{[\mu\bar{\lambda}]}\right)^{\myd}\left(\bar\partial\tilde\eta_{\myd}-\myd\frac{[\mu\hat{\bar\lambda}]}{[\mu\bar{\lambda}]}\frac{\D\hat{\lambda}}{\la\lambda\hat{\lambda}\ra} \tilde\eta_{\myd}+[\kA,\tilde\eta_{\myd}]\right)\,.
\end{equation}
The explicit appearance of arbitrary singular behaviour on the locus $[\mu\bar\lambda]=0$ may seem alarming, but we in fact expect and demand it. For instance, recall that we allow on-shell $H^{0,1}$ Dolbeault representatives which are smooth everywhere on $\mathbb{PT}^{\cO}$ except on the locus $[\mu\bar\lambda]=0$ and $\bar \partial$ closed on $\mathbb{PT}^{\cO}\setminus \{[\mu\bar\lambda]=0\}$. All twistor space conformal primary wavefunctions of the form $\kA=\frac{\Gamma(\Delta-1)\,\bar\delta(\la\lambda\kappa\ra)}{[\mu\tilde\kappa]^{\Delta-1}}$ with $\Delta>1$ are such representatives. These are most naturally interpreted as the pullback of $H^{0,1}$s defined on the twistor space for AdS$_3$ \cite{Bu:2023cef}. After Sparling transforms, their spacetime counterparts will be wavefunctions with poles on $\scri$. 

With this additive field redefinition \eqref{field_redefinition}, the `large gauge transformations' \eqref{PT_lgt_explicit} amounts to $\kA\to \kA'+\sum_{\myd\in\mathbb{Z}_+}\bar\delta^{(\myd-1)}([\mu\bar\lambda])([\mu\hat{\bar\lambda}]\la\lambda\hat{\lambda}\ra)^{\myd}\tilde\eta_{\myd}$. Applying this to $\kB$ as well, with the different gauge parameter $\tilde\phi\in\Omega^0(\mathbb{PT},\cO(-6)\otimes\mathfrak{g})$, we have the following:
\begin{equation}\label{gauge_transformation}
    \kA\to \kA'+\sum_{\myd\in\mathbb{Z}_+}\bar\delta^{(\myd-1)}([\mu\bar\lambda])([\mu\hat{\bar\lambda}]\la\lambda\hat{\lambda}\ra)^{\myd}\tilde\eta_{\myd}\,;\quad \kB\to \kB'+\sum_{\myd\in\mathbb{Z}_+}\bar\delta^{(\myd-1)}([\mu\bar\lambda])([\mu\hat{\bar\lambda}]\la\lambda\hat{\lambda}\ra)^{\myd}\tilde\phi_{\myd}\,,
\end{equation}
where $\kA'$ and $\kB'$ include terms which are not smooth on $\{[\mu\bar\lambda]=0\}$. It is worth noting that \eqref{GT_6d} is only the infinitesimal gauge transformation, a full non-abelian gauge transformation would involve higher order contributions in $\tilde\eta$s. For each order in $\tilde\eta$, we repeat the steps to absorb the pieces in the gauge transformation that do not localise. As for terms that do localise, it suffices that multiple copies of $\bar\delta([\mu\bar\lambda])$ always wedge to $0$, so \eqref{gauge_transformation} stays valid in the full gauge transformations.

After gauge transformation \eqref{gauge_transformation}, the SDYM action on $\mathbb{PT}^{\cO}$ \eqref{QSDYM_6d} stays invariant\footnote{Since there are no codimension $1$ boundary present for $\mathbb{PT}^{\cO}$, hence there are no boundary term under gauge transformations.}. However, with gauge transformations repackaged in \eqref{gauge_transformation}, the action formally separates into two parts:
\begin{multline}
    S_{\text{split}}[\tilde \eta, \tilde \phi,\kA',\kB'] = \int_{\mathbb{PT}^{\cO}} \D^3 Z\wedge \textrm{tr}\left(\kB' \wedge \kF(\kA')\right)-\mathfrak{C}_{\text{YM}}^2\mathrm{Tr}\left(\kA'\partial \kA'\right)\partial\frac{1}{\bar\partial}\mathrm{Tr}\left(\kA'\partial \kA'\right)+\cO(\kA'^5)
    \\+\sum_{\myd}\int_{\mathbb{PT}^{\cO}} \D^3 Z \wedge\bar\delta^{\myd-1}([\mu\bar\lambda])([\mu\hat{\bar\lambda}]\la\lambda\hat{\lambda}\ra)^{\myd}\wedge \textrm{tr}\left(\kB'\wedge\bar\D\tilde\eta_{\myd}+\tilde\phi_{\myd}\bar\D\kA'\right)\,,
\end{multline}
for $\myd\in\mathbb{Z}_+$. The first part is a field redefinition of the original SDYM action \eqref{QSDYM_6d} while the second part localises to formal neighborhoods of the locus $[\mu\bar\lambda]=0$. It is worth noting that the surviving terms from performing transformations \eqref{gauge_transformation} on $S_{\text{ct}}$ \eqref{S_counter} gives the gauge anomaly term which ensures the theory to be free of anomaly while not producing any terms that localise around $\mathbb{CP}^1_{[\mu\bar\lambda]=0}$.

Now we promote $\tilde \eta, \tilde \phi$ to quantum fields (i.e do a path integral over them). This amounts to an enlargement of the gauge symmetries of the theory to include an additive shift symmetry in $\kA', \tilde \eta$ (or $\kB', \tilde \phi$) that preserves the linear combination $\kA$ (or $\kB$).

Notice that because the field redefinition \ref{gauge_transformation} was additive, the Jacobian in the path integral measure is trivial:
\begin{equation}
    \int \D \kA \D \kB = \int \D \kA' \D \kB'\,.
\end{equation}
For correlators involving combinations of $\tilde \eta,\kA'$ and $\tilde \phi,\kB'$ that are in the specific linear combinations $\kA, \kB$, it is always possible to reverse the field redefinitions \ref{gauge_transformation} which decouples the $\tilde \eta, \tilde \phi$ path integrals and recover the original action functional \eqref{QSDYM_6d} with $\kA,\kB$:
\begin{align}
    &\left\la \prod_i \mathcal{O}_i\right\ra_{\text{split}} = \frac{\int \D \kA' \D \kB' \D \tilde \eta \D \tilde \phi \left(\exp\{S_{\text{split}}[\tilde \eta, \tilde \phi,\kA',\kB']\}\prod_i \mathcal{O}_i\right)}{\int \D \kA' \D \kB' \D \tilde \eta \D \tilde \phi \left(\exp\{S_{\text{split}}[\tilde \eta, \tilde \phi,\kA',\kB']\}\right)} \nonumber 
    \\=& \frac{\int \D \kA \D \kB \left(\exp\{S_{\text{split}}[0, 0,\kA,\kB]\}\prod_i \mathcal{O}_i\right)}{\int \D \kA \D \kB \left(\exp\{S_{\text{split}}[0, 0,\kA,\kB]\}\right)} \frac{\text{Vol}(\mathcal{T})}{\text{Vol}(\mathcal{T})}=\frac{\int \D \kA \D \kB \left(\exp\{S_{\text{QSDYM}}\}\prod_i \mathcal{O}_i\right)}{\int \D \kA \D \kB \left(\exp\{S_{\text{QSDYM}}\}\right)}\nonumber 
    \\=&  \left\la \prod_i \mathcal{O}_i\right\ra_{\text{QSDYM}}\,,
\end{align}
where $\mathcal{T}$ is the moduli space of $\tilde\phi$ and $\tilde\eta$ configurations, amounting to the gauge fixing of the additive shift symmetry in which $\tilde\phi$ and $\tilde\eta$ have been gauged away. Therefore the theory which formally split we have written down with the new quantum fields $\tilde \eta, \tilde \phi$ contains QSDYM and can compute all the same correlators QSDYM does (this is admittedly a small set). We can also consider correlators in gauge invariant marginal deformations of QSDYM, $S'=S_{\text{QSDYM}}+\delta S$, and with an identical argument we can show that the split theory for $S'[\tilde\eta,\tilde\phi,\kA',\kB']$ will also be able to compute all the correlators that $S'[0,0,\kA,\kB]$ does. However, in these split theories it is also possible to compute correlators that depend only on $\tilde \eta, \tilde \phi$, which were inaccessible in QSDYM. We will discuss this at length in section \ref{sec:CCFT}.

In the rest of the paper we will specialise to the case with $\myd = 1$ for simplicity of the presentation, we shall discuss potential implications of choosing higher integers $\myd>1$ in the discussion section. The formally split action we consider from now on reads:
\begin{equation}\label{Split_action_PT}
    S_{\text{split}}[\tilde \eta, \tilde \phi] = S_{\text{QSDYM}}[\kA',\kB']
    +\int_{\mathbb{PT}^{\cO}} \D^3 Z \wedge\bar\delta([\mu\bar\lambda])\left([\mu\hat{\bar\lambda}]\la\lambda\hat{\lambda}\ra\right)\wedge \textrm{tr}\left(\kB'\wedge\bar\D\tilde\eta+\tilde\phi\bar\D\kA'\right)\,.
\end{equation}
We would like to stress again that these descriptions of QSDYM can be made equivalent with additive field redefinitions (for correlators that do not depend on fields we wish to decouple).
The additive redefinitions are shift symmetries of the action with trivial path integral Jacobian that decouple fields from the rest of the action. What the $\tilde \eta, \tilde \phi$ variables provide for us is that they bring to the forefront the sub-theory on the singular locus $[\mu\bar\lambda]=0$ of the twistor space large gauge transformations.

\subsection{Geometric significance of $\mathbb{CP}^1_{[\mu\bar\lambda]=0}$}
We would like to understand the importance of the locus $[\mu\bar\lambda]=0$ geometrically. The spinor equation $[\mu\bar\lambda]=0$ has three solutions, namely $\lambda_{\alpha}=0$, $\mu^{\dal}=0$ and $\bar\lambda^{\dal}\propto\mu^{\dal}\neq 0$. We have removed the locus of the first two cases from our definition of $\mathbb{PT}^{\cO}$ already, hence $\mathbb{CP}^1$ locus of $[\mu\bar\lambda]=0$ only has solution when $\bar\lambda^{\dal}\propto\mu^{\dal}\neq 0$. To see the consequence of this on $\mathbb{M}$, we remember the projection between $\mathbb{PT}^{\cO}$ and $\scri^+$ with homogeneous coordinates on the celestial sphere \eqref{scri_metric}:
\begin{equation}
    \sigma: (\mu^{\dal},\lambda_{\alpha})\mapsto (u, \lambda_{\alpha},\bar\lambda_{\dal})\,,
\end{equation}
where $u=[\mu\bar\lambda]=x^{\alpha\dal}\lambda_{\alpha}\bar\lambda_{\dal}$ describes light-cone cuts at $\scri^+$. This projection maps $\mathbb{CP}^1\in\mathbb{PT}^{\cO}$ twistor fibres to celestial spheres on $\scri^+$, where their positions on $\scri^+$ depend on the collective variable $[\mu\bar\lambda]$. Given this projection $\sigma$, our special $\mathbb{CP}^1_{[\mu\bar\lambda]=0}$ can be mapped to the celestial sphere at $u=0$ on $\scri^+$ or equivalently $v=0$ on $\scri^-$. Physically, this means that the class of gauge transformations on $\mathbb{PT}^{\cO}$ using gauge parameters \eqref{eq2} localises a piece of the action functional to the celestial sphere. This precisely matches our intuition of the gauge transformation being `large' on spacetime (non-vanishing on $\scri$). 

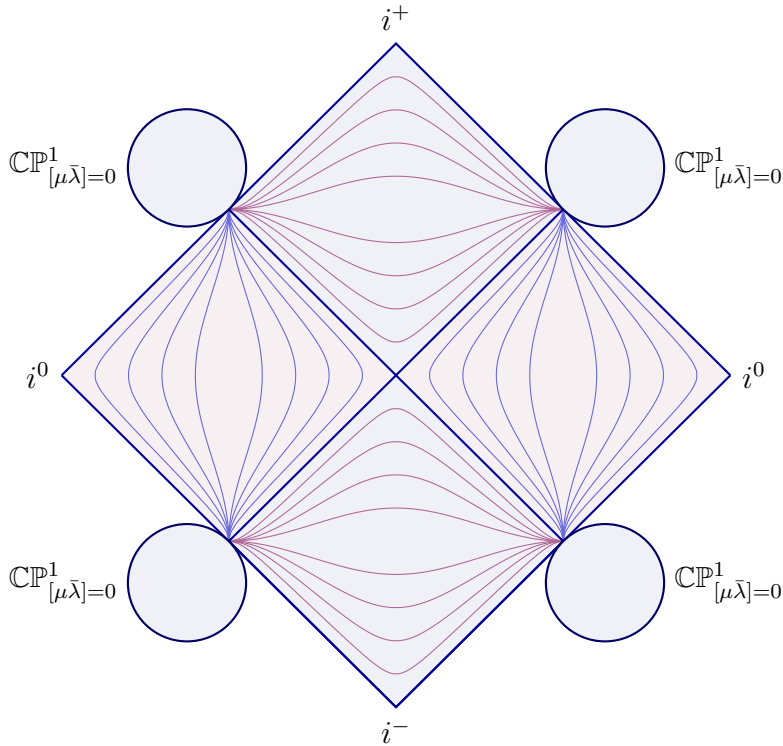
\begin{figure}
    \centering
    \begin{tikzpicture}[scale=2.2]
  \def\Nlines{4} 
  \def\ta{tan(90*1.0/(\Nlines+1))} 
  \def\tb{tan(90*2.0/(\Nlines+1))} 
  \coordinate (O) at ( 0, 0); 
  \coordinate (S) at ( 0,-1); 
  \coordinate (N) at ( 0, 1); 
  \coordinate (W) at (-1, 0); 
  \coordinate (E) at ( 1, 0); 
  \fill[mylightblue] (N) -- (E) -- (S) -- (W) -- cycle;
  \fill[mylightblue] (0,-1) -- (-1,-2) -- (0,-3) -- (1,-2) -- cycle;
  \fill[mylightpurple] (0,-1) -- (1,0) -- (2,-1) -- (1,-2) -- cycle;
  \fill[mylightpurple] (0,-1) -- (-1,0) -- (-2,-1) -- (-1,-2) -- cycle;
  \foreach \i [evaluate={\c=\i/(\Nlines+1); \ct=tan(90*\c);}] in {1,...,\Nlines}{
    \message{  Running i/N=\i/\Nlines, c=\c, tan(90*\c)=\ct...^^J}
    \draw[world line t,samples=\Nsamples,smooth,variable=\t,domain=-1:1] 
      plot(\t,{-penrose(\t*pi/2,\ct)})
      plot(\t,{ penrose(\t*pi/2,\ct)});
  }
    \foreach \i [evaluate={\c=\i/(\Nlines+1); \ct=tan(90*\c);}] in {1,...,\Nlines}{
    \message{  Running i/N=\i/\Nlines, c=\c, tan(90*\c)=\ct...^^J}
    \draw[world line t,samples=\Nsamples,smooth,variable=\t,domain=-1:1] 
      plot(\t,{-penrose(\t*pi/2,\ct)-2})
      plot(\t,{ penrose(\t*pi/2,\ct)-2});
  }
    \foreach \i [evaluate={\c=\i/(\Nlines+1); \ct=tan(90*\c);}] in {1,...,\Nlines}{
    \message{  Running i/N=\i/\Nlines, c=\c, tan(90*\c)=\ct...^^J}
    \draw[world line,samples=\Nsamples,smooth,variable=\x,domain=-1:1] 
      plot({-penrose(\x*pi/2,\ct)+1},\x-1)
      plot({ penrose(\x*pi/2,\ct)+1},\x-1);
  }
      \foreach \i [evaluate={\c=\i/(\Nlines+1); \ct=tan(90*\c);}] in {1,...,\Nlines}{
    \message{  Running i/N=\i/\Nlines, c=\c, tan(90*\c)=\ct...^^J}
    \draw[world line,samples=\Nsamples,smooth,variable=\x,domain=-1:1] 
      plot({-penrose(\x*pi/2,\ct)-1},\x-1)
      plot({ penrose(\x*pi/2,\ct)-1},\x-1);
  }
  \draw[thick, blue!60!black] (0,-1) -- (-1,-2) -- (0,-3) -- (1,-2) -- cycle;
  \draw[thick,blue!60!black] (N) -- (E) -- (S) -- (W) -- cycle;
  \draw[thick,blue!60!black] (-1,0) -- (-2,-1);
  \draw[thick,blue!60!black] (2,-1) -- (1,0);
  \draw[thick,blue!60!black] (0,-3) -- (-2,-1);
  \draw[thick,blue!60!black] (0,-3) -- (2,-1);
  \filldraw[color = mydarkblue!80!black, fill = mylightblue, thick] 
    (1.25,0.25) circle (0.35355);
  \node[right] at (1.6035355,0.25) {$\mathbb{CP}^1_{[\mu\bar\lambda]=0}$};

    \filldraw[color = mydarkblue!80!black, fill = mylightblue, thick] 
    (-1.25,-2.25) circle (0.35355);
    \node[left] at (-1.6035355,-2.25) {$\mathbb{CP}^1_{[\mu\bar\lambda]=0}$};
    
    \filldraw[color = mydarkblue!80!black, fill = mylightblue, thick] 
    (-1.25,0.25) circle (0.35355);
    \node[left] at (-1.6035355,0.25) {$\mathbb{CP}^1_{[\mu\bar\lambda]=0}$};

    \filldraw[color = mydarkblue!80!black, fill = mylightblue, thick] 
    (1.25,-2.25) circle (0.35355);
    \node[right] at (1.6035355,-2.25) {$\mathbb{CP}^1_{[\mu\bar\lambda]=0}$};
    
  \node[above] at (0,1.02) {\large $i^+$};
  \node[below] at (0,-3.01) {\large $i^-$};
  \node[left] at (-2.01,-1) {\large $i^0$};
  \node[right] at (2.01,-1) {\large $i^0$};
\end{tikzpicture}
    \caption{Penrose diagram with Euclidean AdS$_3$ and Lorentzian dS$_3$ slices. $\mathbb{CP}^1_{[\mu\bar\lambda]}=0$ maps to the celestial sphere at the origin light cone cut at $\scri$, which is the common boundary shared by the slices.}
    \label{fig:penrose_diagram}
\end{figure}

Scaling reductions of twistor space $\mathbb{PT}^{\cO}$ along complexified null geodesics in the corresponding complexified Minkowski space give minitwistor spaces $\mathbb{MT}:=\{\mathbb{CP}_{\lambda}^1\times\mathbb{CP}_{\mu}^1\,|\,[\mu\bar\lambda]\neq 0\}$ of EAdS$_3$ or LdS$_3$ \cite{Jones:1985pla,LeBrun:2008ch}, which are slices of 4d Minkowski space in the Milne and Rindler regions (given explicit choice of light-cone at the origin). Note that the sphere of discussion here is excluded from the definition of $\mathbb{MT}$. The light cone of the origin divides Minkowski space into regions, where an associated minitwistor space exists for each region. The celestial sphere $\mathbb{CP}^1_{[\mu\bar\lambda]=0}$ at $u=0$ on $\scri^+$ or $v=0$ on $\scri^-$ happens to be the common boundary sphere for all the EAdS$_3$ or LdS$_3$ slices as shown in figure \ref{fig:penrose_diagram}. This indicates that the partial localisation of the SDYM action under special large gauge transformations can be interpreted holographically in 3d.

\section{Scaling reduction and 3d splitting}\label{sec:3d}
In \cite{Bu:2023cef}, the authors have investigated the $\mathbb{C}^*$ scaling reduction of 4d conformal theories to hyperbolic slices of complexified Minkowski space using the geometric machinery of (mini-)twistor spaces. Using similar methods here, we can $\mathbb{C}^*$ scaling reduce the first part of the twistor action \eqref{Split_action_PT} to minitwistor spaces for Milne/Rindler regions depending on the patch choice $x^2>0$ or $x^2<0$, whereas the second part localises to $\mathbb{CP}^1_{[\mu\bar\lambda]=0}$ (which can be mapped to the boundary celestial sphere covering $x^2=0$). Beginning with complexified Minkowski space $\mathbb{M}_{\mathbb{C}}$, the embedding space metric can be written as 
\begin{equation}\label{embed_metric}
    \d s^2_{\text{3d}}=\frac{1}{x^2}\left(\d x^2-\frac{(x\cdot\d x)}{x^2}\right)\,,
\end{equation}
where $x^{\alpha\dal}$ are coordinates in 4d. One can choose to focus on the patch where $x^2>0$ or $x^2<0$, which amounts to picking real or imaginary radius satisfying $x^2=r^2$. This gives the embedding of EAdS$_3$ or LdS$_3$ slices in Minkowski space. Note that given the form of the embedding metric \eqref{embed_metric}, one is allowed to rescale $x^{\alpha\dal}$ by arbitrary non-zero function $f(x)$ since the metric is basic and invariant under the Euler vector field $\Upsilon=x^{\alpha\dal}\partial_{\alpha\dal}$. Geometrically, this essentially allows us to consider continuous deformations generated by position dependent $x$ rescalings of embedded EAdS$_3$ or LdS$_3$ slices with their boundary 2-spheres lying on the projectivised null cone of the origin.

\paragraph{Scaling mode decomposition on the locus $\{[\mu \bar \lambda]\neq 0\}\subset\mathbb{PT}^{\cO}$}:
The locus $\{[\mu \bar \lambda]\neq 0\}\subset\mathbb{PT}^{\cO}$ is the total space of a topologically trivial $\mathcal{O}(1,-1)$ bundle over $\mathbb{MT}$ (with the zero section removed). By decomposing our fields in a suitable basis on the fibre, we can perform the integral over the fibre coordinate and trade the fibre degrees of freedom for a complex mode number.

The first term in \eqref{Split_action_PT} gives the minitwistor action as derived in \cite{Bu:2023cef}:
\begin{multline}
   \int\d^2\Delta \int\d^2\Delta' \, \int_{\mathbb{MT}} \D\lambda\wedge\D\mu\wedge\left(\tr\left(\eta_{\Delta}\bar\partial a_{-\Delta}+b_{\Delta}\wedge\bar\partial\phi_{-\Delta}+(2\im-\Delta) b_\Delta \wedge a_{-\Delta}\right) \right.
   \\\left.+ \,\tr\left(\eta_{\Delta+\Delta'} a_{-\Delta}\wedge a_{-\Delta'}+b_{\Delta+\Delta'}\wedge a_{-\Delta} \phi_{-\Delta'}\right)+S_{\text{ct-3d}}\right)\,,
\end{multline}
where $\Delta=m+\im\rho\in\mathbb{Z}+\im\mathbb{R}$ labels the complex mode number from mode decompositions in the $\mathbb{C}^*$ scaling direction. More precisely, $\mathbb{C}^*\cong S^1\times\mathbb{R}_+$, the real part $m\in\mathbb{Z}$ is the single valued discrete $S^1$ Fourier mode number, the imaginary part $\rho\in\mathbb{R}$ is the usual Mellin mode number $1+\im\rho$ on the principal series representation after a Mellin transform in $\mathbb{R}_+$. The integral sign over $\Delta=m+\im\rho$ is only formal, we use it do denote $\sum_{m\in\mathbb{Z}}\int_{\rho\in\mathbb{R}_+}\d\rho$.
There is a small subtlety we have glossed over in the mass term in the first line of the equation above, the offset $2 \im$ is only present for one of the two terms in $b\wedge a$. The explicit detail can be found in appendix \ref{appendix:detail_scaling_reduction}\footnote{Note that although in principle our formalism allows for arbitrary $\Delta \in \mathbb{C}$, or $m,\rho\in\mathbb{R}$, mode coefficients for fields that came from a single valued field on $\mathbb{PT}^{\cO}$ will only have nonvanishing mode coefficients when the $S^1$ modes are discrete.}. 

$a_{\Delta}$ and $b_{\Delta}$ are the Mellin transformed gluons (on-shell) with independent homogeneous scalings in $\lambda_{\alpha}$ and $\mu^{\dal}$, we label their individual scalings weights in the following way: $b_{\Delta}\in\Omega^{0,1}\left(\mathbb{MT},\mathcal{O}_{\Delta-3}(-4)\otimes\frak{g}\right)$, $a_{\Delta}\in\Omega^{0,1}\left(\mathbb{MT},\mathcal{O}_{\Delta-1}(0)\otimes\frak{g}\right)$. As discussed in Appendix $\ref{appendix:detail_scaling_reduction}$, the scaling weight in each $\mathbb{CP}^1$ factor of $\mathbb{MT}$ can be read off from the real part of the subscript of $\cO$, while the integer number in the bracket denotes the sum of the scaling weights in $\lambda_{\alpha}$ and $\mu^{\dal}$. $\phi$ and $\eta$ denote the $\kA'$ and $\kB'$ pointing in the direction of the scaling reduction in $\mathbb{PT}^{\cO}$, which become scalars on $\mathbb{MT}$, $\phi\in\Omega^0\left(\mathbb{MT},\mathcal{O}_{-\Delta+1}(0)\otimes\frak{g}\right)$, $\eta\in\Omega^0\left(\mathbb{MT},\mathcal{O}_{-\Delta-1}(-4)\otimes\frak{g}\right)$. $a_{\Delta}$ and $b_{\Delta}$ are sections of line bundles on $\mathbb{MT}$ labelled by  $\Delta\in\mathbb{Z}+\im\mathbb{R}$ and spin $s\in\mathbb{Z}$. On EAdS$_3$, they correspond on-shell to fields satisfying the Laplace equation with mass controlled by $\Delta$. Detail of this has been worked out in \cite{Bailey_Dunne_1998,Tsai1996ThePT,Bu:2023cef}. $S_{\text{ct-3d}}$ denotes the scaling reduced counter term \eqref{S_counter} for the gauge anomaly on $\mathbb{PT}^{\cO}$. Its explicit form is obscure and complicated so we do not reproduce it here. The important thing is that it and the rest of the action are literally equal to the anomaly-free theory on $\mathbb{PT}^{\cO}$, as we have kept track of all the modes in the decomposition.

\paragraph{Scaling mode decomposition on the locus $\{[\mu \bar \lambda] = 0\}\subset\mathbb{PT}^{\cO}$:}
The locus $\{[\mu \bar \lambda]= 0\}\subset\mathbb{PT}^{\cO}$ is the total space of a topologically trivial $\mathcal{O}(2)$ bundle over $\mathbb{CP}^1_{\lambda}$ (with the zero section removed). On the locus $\{[\mu \bar \lambda]=0\}\subset\mathbb{PT}^{\cO}$, we may decompose every appearance of $\mu^{\dot \alpha}$ in the following way:
\begin{align}
    \mu^{\dot \alpha} = \epsilon^{\dot \alpha \dot \beta} \mu_{\dot \beta} = \left(\frac{\bar \lambda^{\dot \alpha}\hat{\bar \lambda}^{\dot \beta}-\hat{\bar \lambda}^{\dot \alpha}\bar \lambda^{\dot \beta}}{\la \lambda \hat \lambda \ra}\right)\mu_{\dot \beta} = \frac{[\mu\bar \lambda]}{\la \lambda \hat \lambda \ra} \hat{\bar \lambda}^{\dot \alpha} - \frac{[\mu \hat{\bar \lambda}]}{\la \lambda \hat \lambda \ra} \bar \lambda^{\dot \alpha} = 0 - \frac{[\mu \hat{\bar \lambda}]}{\la \lambda \hat \lambda \ra} \bar \lambda^{\dot \alpha}
\end{align}
On restriction to the locus $[\mu \bar \lambda]=0$, the only $\mu$ dependence is therefore on $[\mu \hat{\bar \lambda}]\in \mathcal{O}(2)$ and on its complex conjugate. The decomposition into eigenstates under the $\mathbb{C}^*$ scaling of $\mu^{\dot \alpha}$ and the conjugate action on $\bar \mu^{\alpha}$ (in which the factors of $\la \lambda \hat \lambda \ra$ are included for convenience to balance antiholomorphic weight\footnote{This allows us to stop keeping track of the antiholomorphic scaling weight, it is always fixed to be $0$.}) gives:
\begin{multline}
    \left.\kA'(\mu^{\dot \alpha}, \lambda^{\alpha})\right\vert_{[\mu \bar \lambda]=0} = \kA'\left(- \frac{[\mu \hat{\bar \lambda}]}{\la \lambda \hat \lambda \ra} \bar \lambda^{\dot \alpha}, \lambda^{\alpha} \right) \\
    = \int_{\mathbb{R}} \d\rho \sum_{m\in\mathbb{Z}} \left([\mu \hat{\bar \lambda}]\right)^{(m+\im \rho)/2}\frac{\left( [\hat \mu \bar{\lambda}]\right)^{(-m+\im \rho)/2}}{\la \lambda \hat \lambda \ra^{-m+\im \rho}} a'_{m+\im \rho}(\lambda^{\alpha},\bar\lambda^{\dal}) \,,
\end{multline}
where we have expanded $\kA'$ in the $\cO(2)$ fibre coordinates. The remaining modes $a'_{m+\im\rho}(\lambda^{\alpha},\bar\lambda^{\dal})$ lives on $\mathbb{CP}^1_{[\mu\bar\lambda]=0}$:
\begin{equation}\label{a_restriction}
    \boxed{a'_{m+\im\rho}(\lambda^{\alpha}, \bar \lambda^{\dot \alpha}) \in \Omega^{0,1}(\mathbb{CP}^1_{[\mu \bar \lambda]=0}, \mathcal{O}(-2m)\otimes \mathfrak{g})}\,.
\end{equation}
Interested readers are referred to Appendix \ref{appendix:detail_scaling_reduction} for full details. Decomposing each field into scaling eigenvalues and then integrating out the fibre means that the second term of \eqref{Split_action_PT} becomes:
\begin{multline}\label{localised_part_3d}
\int\d^2\Delta\int\d^2\Delta'\,\int_{\mathbb{CP}^1_{[\mu\bar\lambda]=0}}\D\lambda\wedge\tr\left(-\eta_{\Delta}\bar\partial\tilde\eta_{-\Delta}+\tilde\phi_{\Delta}\bar\partial\phi_{-\Delta}-\Delta \tilde\phi_{\Delta} a'_{-\Delta}-(\Delta-2\im) b_{\Delta}\tilde \eta_{-\Delta}+\right.
\\\left. \eta_{\Delta+\Delta'}[\tilde\eta_{-\Delta},a'_{-\Delta'}]+\tilde\phi_{\Delta+\Delta'}\phi_{-\Delta} a'_{-\Delta'}+\phi_{\Delta+\Delta'}\tilde\eta_{-\Delta} b'_{-\Delta'}\right)\,,
\end{multline}
where we have performed the integral over $\mu$ against the delta function $\bar\delta([\mu\bar\lambda])$\footnote{If we were also considering other positive integers $\gamma>1$, one needs to integrate by parts on the normal derivatives $\bar\delta^{(\gamma-1)}([\mu\bar\lambda])$ pointing off of the $\mathbb{CP}^1$ and distribute them on the remaining part of the integrand. This would select fields that have higher order dependence on $[\mu\bar\lambda]$, which are extendable to higher order formal neighborhoods of $\mathbb{CP}^1_{[\mu\bar\lambda]}$.}. $a'$ and $b'$ denote the gluon representatives $a$ and $b$ when localised to $\mathbb{CP}^1_{[\mu\bar\lambda]=0}$.

\medskip

As we mentioned $\Delta\in\mathbb{Z}+\im\mathbb{R}$, it seems that we are declaring $a'$, $b'$ are sections of a line bundle labelled by a complex number and an integer\footnote{The Birkhoff-Grothendieck theorem \cite{Birkhoff1909,Grothendieck1957} states that holomorphic bundles over projective line $\mathbb{CP}^1$ are direct sums of holomorphic line bundles, which have integer labels.}. Although we have demonstrated that the choice of $\Delta$ does not affect the scalings (degree of the line bundle) \eqref{a_restriction}, it is interesting to point out that such bundles have appeared in \cite{Held:1970kr,Newman:1966ub}. The authors attempted to describe asymptotic fields with conformal weight $\Delta\in\mathbb{Z}+\im\mathbb{R}$ and spin $s\in\mathbb{Z}$ on the celestial sphere. It was further established in \cite{Curtis_saver}, we shall briefly describe the construction on $\mathbb{CP}^1$, interested readers are directed to their original articles. 

To construct a line bundle with labels $(\delta,s)$, we begin with the principle bundle $\mathbb{C}^2\setminus\{0\}\xrightarrow{\pi}\mathbb{CP}^1$ with the fibres being $\mathbb{C}^*$. Pick a point $p\in\mathbb{B}\cong\mathbb{C}^2\setminus\{0\}$, $(\pi(p)=\lambda,z)$ with $z\in\mathbb{C}$ gives a point on $\mathbb{CP}^1$ together with a point on tangent space. Consider the equivalence class of such $\{p,z\}$ under $\mathbb{C}^*$: $(p,z)\sim (pg,\rho(g^{-1})z)$. $g$ denotes operation $\mathbb{C}^*$ on $p$ while $\rho(g^{-1})$ denotes the representation of $\mathbb{C}^*$ operation on $\mathbb{C}$. Since $(\delta,s)$ defines a mapping $\lambda\to \lambda^{-s-\delta}\bar\lambda^{-s+\delta}$, which gives a representation $\rho$ of $\mathbb{C}^*$ on $\mathbb{C}$, this equivalence class is a trivial bundle itself: $\mathbb{CP}^1\times\mathbb{C}\to\mathbb{CP}^1$. It is natural to think about the functions on $\mathbb{CP}^1$ with conformal weight $\Delta$ and spin $s$ as sections of this bundle. Its transition function can be defined through the transition function of the principle bundle after local trivialisation, explicit demonstration is suppressed here to avoid further diversion. This construction seems to contradict the fact that Chern classes characterising line bundles on $\mathbb{CP}^1$ are integers: $H^2(\mathbb{CP}^1,\mathbb{Z})\cong\mathbb{Z}$. As mentioned in \cite{Curtis_saver}, our bundle is in fact homotopic to one with $\delta=0$, which has integer Chern class. 

\medskip

\paragraph{The complete scaling reduced action}:
After fixing the detail of the action functional localising to $\mathbb{CP}^1_{[\mu\bar\lambda]=0}$, we collect all the terms and write down the scaling reduced actions (suppressing the $\Delta$ labels):
\begin{multline}\label{MT_split_action}
  S_{\text{split-3d}}=
  \\ \int_{\mathbb{Z}+\im\mathbb{R}}\d^2\Delta\,  \int_{\mathbb{MT}} \D\lambda\wedge\D\mu \,\tr\left(\eta\bar\partial a+\eta a\wedge a+b\wedge\bar\partial\phi+b\wedge a \phi-\Delta b \wedge a-2\im (b \wedge a)\right)+S_{\text{ct-3d}}\\ 
  +\int_{\mathbb{Z}+\im\mathbb{R}}\d^2\Delta\,\int_{\mathbb{CP}^1_{[\mu\bar\lambda]=0}}\D\lambda\wedge\tr\left(-\eta\bar\partial\tilde\eta+\eta[\tilde\eta,a']+\tilde\phi\bar\partial\phi+\tilde\phi\phi a'+\phi\tilde\eta b'-\Delta \tilde\phi a'-(\Delta-2 \im) b\tilde\eta\right)\,.
\end{multline}
One other comment we would like to make is that the meaning of the particular gauge parameters \eqref{PT_1_gold} is clear in $\mathbb{M}$ as large gauge parameters \eqref{eq3}. After scaling reductions, they have a natural interpretation as spin $1$, $\Delta=1$ bulk-to-boundary propagators on EAdS$_3$ slices of Minkowski space. With the simplest $\tilde\eta$ (having trivial $\mu$ dependence), we see that 
\begin{equation}
     A_{\mu} dx^{\mu}= d\left(\frac{x\cdot\veps}{x\cdot p}\right)=\frac{\veps_{[\mu}p_{\nu]}\,x^{\mu}dx^{\nu}}{(x\cdot p)^2} = \frac{\kappa^{(\alpha}\kappa^{\beta)}\,[\tilde\kappa\hat{\tilde\kappa}]\,x^{\dal}_{\alpha}dx_{\dal \beta}}{(\la\kappa| x |\tilde \kappa])^2}\,,
\end{equation}
where $\la\kappa| x |\bar \kappa]=x^{\alpha\dal}\kappa_{\alpha}\tilde\kappa_{\dal}$.
This agrees with the known form for the bulk-to-boundary propagator \cite{Costa:2014kfa} and can be checked to be invariant and basic under the Euler vector field. Generic $\tilde \eta(\lambda)\rightarrow \tilde \eta(\kappa)$ amounts to allowing a convolution over $\kappa$ to generate more generic profiles for large gauge transformations in 4d, or equivalently, generic on-shell bulk field configurations in 3d. The (de Rham $d$-exact) 1-form on $\mathbb{M}$ is basic and invariant under the Euler vector field and therefore is pulled back from EAdS$_3$/LdS$_3$, depending on the value of $x^2$.

\section{2d current algebra}\label{sec:CCFT}
Since we have kept track of all the modes $\Delta\in\mathbb{Z}+\im\mathbb{R}$, the scaling reduced formally split action \eqref{MT_split_action} is still equivalent to the original QSDYM action \ref{QSDYM_6d} on $\mathbb{PT}^{\cO}$. We have already studied the minitwistor action on $\mathbb{MT}$ and its relation to EAdS$_3$ explicitly in \cite{Bu:2023cef}. Given the purpose for this paper is to derive a 2d CFT producing dynamics of SDYM using spacetime large gauge transformations, we focus on the system which only lives on $\mathbb{CP}^1_{[\mu\bar\lambda]=0}$ \eqref{localised_part_3d} and recognize it as a 2d chiral CFT with infinitely many fields and conformal primaries labelled by $\Delta\in\mathbb{Z}+\im\mathbb{R}$. In fact, we shall see that in section \ref{sec_ff_cf}, in the computation of correlation functions on $\mathbb{CP}^1_{[\mu \bar \lambda]=0}$, the $\mathbb{MT}$ action completely decouples. The correlators producing space time Yang-Mills form factors rely solely on the 2d CFT. Zooming in on the part of the action (suppressing the labels on the fields):
\begin{equation}
   \int_{\mathbb{Z}+\im\mathbb{R}}\d^2\Delta\, \int_{\mathbb{CP}^1_{[\mu\bar\lambda]=0}}\tr\left(-\eta\bar\partial\tilde\eta+\eta[\tilde\eta,a']+\tilde\phi\bar\partial\phi+\tilde\phi\phi a'+\phi\tilde\eta b'+\Delta \tilde \phi a'-(\Delta-2\im) b \tilde \eta\right)\,,
\end{equation}
where we have absorbed the top form $\D\lambda$ into the definitions of $\tilde\eta$ and $\tilde\phi$. This can be seen as a chiral CFT with infinitely many fields labeled by $\Delta\in\mathbb{Z}+\im\mathbb{R}$ defining a chiral algebra and its coupling to gluon modes $a'$ and $b'$. Define $J, \tilde J$ as the currents that couple to $a'$, $b'$:
\begin{align}\label{def_generator}
    &J^{\msf{a}}_{\Delta}=\int\d^2j\, \f^{\msf{a}}{}_{\msf{mn}}\left(:\eta^{\msf{m}}_{j}\tilde\eta^{\msf{n}}_{\Delta-j}: +:\phi^{\msf{m}}_{j}\tilde\phi^{\msf{n}}_{\Delta-j}:\right)-\Delta\tilde \phi_{\Delta}^{\msf{a}}\,,
    \\
    &\tilde J^{\msf{a}}_{\Delta}=\int\d^2j\, \f^{\msf{a}}{}_{\msf{mn}}\left(:\phi^{\msf{m}}_{j}\tilde\eta^{\msf{n}}_{\Delta-j}:\right)+(\Delta-2\im) \tilde \eta^{\msf{a}}_{\Delta}\,,
\end{align}
where $\f^{\msf{a}}_{\msf{mn}}$ denotes the structure constant of the gauge group. This allows us to rewrite the action simply as (suppressing the $\Delta$ labels):
\begin{equation}\label{P1_current_action}
    S_{\text{current}}=\int \d^2\Delta\int_{\mathbb{CP}^1_{[\mu\bar\lambda]=0}}\tr\left(-\eta\bar\partial\tilde\eta+a' J+\tilde\phi\bar\partial\phi+b'\tilde J\right)\,.
\end{equation}
In order for the action to make sense in homogeneity in $\lambda_\alpha$, one could first recall that $a'_{\Delta=m+\im \rho}\in\Omega^{0,1}\left(\mathbb{CP}^1_{[\mu\bar\lambda]=0},\cO(-2m)\right)$, which dictates the homogeneity of the current to be $J_{-\Delta=-m+-\im \rho}^{\msf{a}}\in\Omega^{1,0}\left(\mathbb{CP}^1_{[\mu\bar\lambda]=0},\cO(2m)\right)$. Similarly, the kinetic terms determines the fact that there are only propagators between $\eta$ and $\tilde\eta$ when their modes sum to $0$. 

\subsection{Operator product expansions}
In this section, we would like to further compute the operator product expansion (OPE) between current algebra generators $J^{\msf{a}}$ and $\tilde J^{\msf{a}}$. Their OPEs are given by the chiral propagator between the component fields:
\begin{equation}
    \eta^{\msf{a}}_{\Delta_1}(z_1,\bar z_1)\tilde\eta^{\msf{b}}_{\Delta_2}(z_2,\bar z_2) \sim \frac{\kappa^{\msf{ab}}\,\d z_2}{z_{12}}\,\delta_{\Delta_1+\Delta_2,0}\,,
\end{equation} 
where we have used local coordinates $z,\bz$ on $\mathbb{CP}^1$. The mode number matching delta function $\delta_{\Delta_1+\Delta_2,0}$ comes from the requirement for the $\bar\partial$ kinetic terms in \eqref{P1_current_action} to be well defined. Similar OPE exists between $\phi$ and $\tilde\phi$, while there are no other possible OPEs. The generator $J^{\msf{a}}_{\Delta}$ with mode number $\Delta$ defined in \eqref{def_generator} is explicitly:
\begin{equation}
    J^{\msf{a}}_{\Delta}=\int\d^2j \f^{\msf{a}}{}_{\msf{mn}}\left(:\eta^{\msf{m}}_{j}\tilde\eta^{\msf{n}}_{\Delta-j}: +:\phi^{\msf{m}}_{j}\tilde\phi^{\msf{n}}_{\Delta-j}:\right)-\Delta\tilde\phi_{\Delta}^{\msf{a}}\,,
\end{equation}
where the sum over $j$ is summing over all possible weight of the first component field. With these, we can compute the OPE between the generators, for simplicity, we have suppressed the $\phi\tilde\phi$ terms which works in the same fashion.
\begin{multline}
    J^{\msf{a}}_{\Delta_1}(z_1,\bz_1)J^{\msf{b}}_{\Delta_2}(z_2,\bz_2) \sim \int\d^2j\,\d^2l \,\f^{\msf{a}}_{\msf{mn}}\f^{\msf{b}}_{\msf{pq}}\times\left(\frac{\delta^{\msf{mq}}\delta^{\msf{np}}\delta_{\Delta_1-j+l,0}\delta_{\Delta_2+j-l,0}\d z_1\d z_2}{z_{12}^2}\,\right.\\\left.+\tilde\eta^{\msf{n}}_{\Delta_1-j}\eta^{\msf{p}}_{l}\,\frac{\delta^{\msf{mq}}\delta_{j+\Delta_2-l,0}\d z_1}{z_{12}}-\eta^{\msf{m}}_{j}\tilde\eta^{\msf{q}}_{\Delta_2-l}\,\frac{\delta^{\msf{np}}\,\delta_{\Delta_1-j+l,0}\d z_2}{z_{12}}\,+\phi\tilde\phi\text{ terms}\right)
    \\
    +\f^{\msf{a}}_{\msf{mn}} \kappa^{\msf{nb}} \tilde \phi^{\msf{m}}_{\Delta_1+\Delta_2}\frac{\Delta_1+\Delta_2}{z_{12}}
    \,.
\end{multline}
In the OPE limit, both terms are evaluated at $(z_2,\bz_2)$, after eliminating the sum over $l$ with the delta functions:
\begin{multline}
    J^{\msf{a}}_{\Delta_1}(z_1,\bz_1)J^{\msf{b}}_{\Delta_2}(z_2,\bz_2) \sim \int\d^2j\,\frac{\kappa^{\msf{ab}}\delta_{\Delta_1+\Delta_2,0}\d z_1\d z_2}{z_{12}^2}\,+\\\f^{\msf{aq}}_{\msf{n}}\f^{\msf{b}}_{\msf{pq}}\,\tilde\eta^{\msf{n}}_{\Delta_1-j}\eta^{\msf{p}}_{j+\Delta_2}\,\frac{\d z_2}{z_{12}}-\f^{\msf{ap}}_{\msf{m}}\f^{\msf{b}}_{\msf{pq}}\,\eta^{\msf{m}}_{j}\tilde\eta^{\msf{q}}_{\Delta_2+\Delta_1-j}\,\frac{\d z_2}{z_{12}}\,+\phi\tilde\phi\text{ terms}\\
    +\f^{\msf{a}}_{\msf{mn}} \kappa^{\msf{nb}} \tilde\phi^{\msf{m}}_{\Delta_1+\Delta_2}\frac{\Delta_1+\Delta_2}{z_{12}}\,,
\end{multline}
where $\kappa^{\msf{ab}}$ is the Killing form of the Lie algebra $\mathfrak{g}$. Notice that we have used the weight delta functions in the simple pole terms to eliminate the sum over $l$, while the two delta functions in the double pole term combined to one, the remaining sum over $j\in\mathbb{Z}_+$ would produce infinitely many such double poles. After appropriately adjusting the dummy adjoint indices and shifting $j\to (\Delta_1-j)$ in the second term, we see that the two terms on the second line combines. Summarizing this, we have
\begin{equation}\label{JJ_OPE}
    J^{\msf{a}}_{\Delta_1}(z_1,\bz_1)J^{\msf{b}}_{\Delta_2}(z_2,\bz_2) \sim \frac{\f^{\msf{ab}}{}_{\msf{c}}\,J^{\msf{c}}_{\Delta_1+\Delta_2}(z_2,\bz_2)}{z_{12}}+ \infty\times \frac{\kappa^{\msf{ab}}\,\delta_{\Delta_1+\Delta_2,0}\,\d z_1\d z_2}{z_{12}^2}\,,  
\end{equation}
where the $\infty$ sign in front of the double pole term formally denotes the fact that infinitely many double pole terms are present. This formally mimics the structure of a Kac-Moody algebra with the level $k=0$ or infinitely many copies with level $k=1$ depending on the delta function $\delta_{\Delta_1+\Delta_2,0}$, which indicates that the double pole terms only exist if $\Delta_1+\Delta_2=0$. Physically, this corresponds to when the modes $a'$ that couple to the currents satisfy $\Delta_1\leq 0\leq\Delta_2$ or the other way round. For all other choices of $\Delta$s, the double pole terms are absent. Importantly, this suggests that for most of the gluon modes, our corresponding chiral CFT is \textit{non-unitary}. For the rest of the paper, we shall focus on the subsector of the OPEs where the two current algebra generators have their mode numbers $\Delta_1+\Delta_2\neq 0$ excluding the existence of the double pole. 

For a 2d current algebra valued in the Lie algebra $\mathfrak{g}$, quantum mechanically, its central charge is related to the level $k$ of the current algebra and the dimension of $\mathfrak{g}$ \cite{Polchinski:1998rr}:
\begin{equation}
    c= \frac{k\,\text{dim}(\mathfrak{g})}{k+h^{\vee}}\,,
\end{equation}
where $h^{\vee}$ is the dual Coxeter number.  We shall derive the regularised expression for $c$ by explicitly calculating the fourth order pole in the stress tensor self-OPE and show that with the removal of the regulator, $c \rightarrow \infty$:
\begin{equation}
    c = \lim_{\Omega\rightarrow \infty}(\pi\Omega^2)\frac{k(\Omega)\,\text{dim}(\mathfrak{g})}{k(\Omega)+h^{\vee}}\to\infty\,,
\end{equation}
where $\Omega\in\mathbb{R}^+$ denotes a finite truncation of the spectrum of $\Delta$ and $k(\Omega)$ is the level evaluated on the truncated spectrum. Similarly, one can work out OPEs between $J^{\msf{a}}$ and $\tilde J^{\msf{b}}$ using the component field propagators. And the OPEs between the $\tilde J^{\msf{a}}$ generators \eqref{def_generator} are regular given that there are no available propagators between $\phi$ and $\tilde\eta$. Summarizing all the OPEs between the generators:
\begin{equation}\label{S-algebra}
    \begin{cases}
    J^{\msf{a}}_{\Delta_1}(z_1,\bz_1)J^{\msf{b}}_{\Delta_2}(z_2,\bz_2) \sim \frac{\f^{\msf{ab}}{}_{\msf{c}}\,J^{\msf{c}}_{\Delta_1+\Delta_2}(z_2,\bz_2)}{z_{12}}+ \infty\times \frac{\kappa^{\msf{ab}}\,\delta_{\Delta_1+\Delta_2,0}\,\d z_1\d z_2}{z_{12}^2}\,;\\[10pt]
    J^{\msf{a}}_{\Delta_1}(z_1,\bz_1)\tilde J^{\msf{b}}_{\Delta_2}(z_2,\bz_2) \sim \frac{\f^{\msf{ab}}{}_{\msf{c}}\,\tilde J^{\msf{c}}_{\Delta_1+\Delta_2}(z_2,\bz_2)}{z_{12}}\,;\\[10pt]
    \tilde J^{\msf{a}}_{\Delta_1}(z_1,\bz_1)\tilde J^{\msf{b}}_{\Delta_2}(z_2,\bz_2) \sim 0 \,.
    \end{cases}
\end{equation}

\medskip

This particular set of chiral algebras is the infinite dimensional symmetry algebra has appeared in the literature with various derivations for $\Delta\in\{1,0,-1,-2,\dots\}$, which is a subset of our label $\Delta\in\mathbb{Z}+\im\mathbb{R}$. This is demonstrating the fact that we are allowing off-shell degree of freedom in the action while the S-algebra uses generators that come from on-shell scattering states. The on-shell derivations have been very well developed in the literature, \cite{Costello:2022wso} used Koszul duality which requires gauge invariant couplings between the 2d currents and bulk SDYM fields. \cite{Guevara:2021abz} expanded Mellin transformed holomorphic collinear splitting functions in conformally soft modes and \cite{Adamo:2021zpw} derived it from ambitwistor string worldsheet theories. Usually referred to as the S-algebra, \eqref{S-algebra} underlies the symmetry of the self-dual sector of Yang-Mills theory realized as the loop algebra of endomorphisms of the gauge bundle $E$ over complex null 2-planes $\mathbb{C}^2$: $\mathcal{L}\text{End}E(\mathbb{C}^2)$. The statement we are making here is that, one can derive it on the level of the action by isolating the large gauge transformations with localising degrees of freedom among the rest of the field configurations.

\subsection{Stress tensor}
In order to work with the $\Delta$-space volume factor $\textrm{Vol}(\mathbb{Z}+\im\mathbb{R}) = \infty$, define a $\Omega$-\textit{truncation} to be setting
\begin{equation}
    (\eta_{\Delta},\tilde \eta_{\Delta}, \phi_{\Delta}, \tilde \phi_{\Delta}) = 0 \quad \textrm{if }\Delta \notin D(\Omega) \subset \mathbb{Z}+\im\mathbb{R}\,,
\end{equation}
in which $D(\Omega):= \{|z|\leq \Omega\}$ is the disc of radius $\Omega \in \mathbb{R}^+$ centred at the origin. We will work with large $\Omega$ throughout and then take it to $\infty$ at the end of the computations. Under the $\Omega$ truncation, we have the level of the current algebra as $k(\Omega)=\textrm{Vol}(D(\Omega)) \approx \pi \Omega^2$. Consider the weighted stress tensor of the chiral current algebra system \eqref{P1_current_action}:
\begin{equation}
    H_{\Delta} = \frac{1}{2k(\Omega)+C_2}\int_{\text{Vol}(D(\Omega))}\d^2q\sum_{\msf{a}}:\tilde\eta^{\msf{a}}_{q}\,\partial\eta^{\msf{a}}_{\Delta-q}+\tilde\phi^{\msf{a}}_{q}\,\partial\phi^{\msf{a}}_{\Delta-q}:\,,
\end{equation}
where the sum runs over both the adjoint index $\msf{a}$ and scaling weight $q$. $C_2$ denotes the quadratic Casimir of the adjoint representation of the gauge group $G$. The OPE between $H_{\Delta_i}$ and $H_{\Delta_j}$ reads
\begin{equation}
    H_{\Delta_i}(z_1)H_{\Delta_j}(z_2) \sim \frac{\delta_{\Delta_i+\Delta_j,0}\,c/2}{z_{12}^4} + \frac{H_{\Delta_{i}+\Delta_j}}{z_{12}^2}+ \frac{\partial H_{\Delta_i+\Delta_j}}{z_{12}}\,,
\end{equation}
where the weight-matching delta function $\delta_{\Delta_i+\Delta_j,0}$ is the counterpart of the condition for the level term to exist in the $JJ$ OPE \eqref{JJ_OPE}. We see that for $H_{\Delta}$ to be the stress tensor in the 2d CFT \eqref{P1_current_action}, we need to take the zero mode in the scaling reduction $\Delta=0$. This is natural from the upstairs theory, SDYM is conformal field theory on $\mathbb{PT}^{\cO}$, hence $H_{\Delta}$ with all the modes summed over served as a stress tensor for the 6d CFT. Only the zero mode $H_0$ genuinely lives on $\mathbb{CP}^1_{[\mu\bar\lambda]=0}$. We shall refer to $T=H_0$ as the stress-energy tensor of our 2d CFT.
\begin{equation}\label{TT_OPE}
    T(z_1)T(z_2) \sim \frac{c/2}{z_{12}^4} + \frac{T}{z_{12}^2}+ \frac{\partial T}{z_{12}}\,.
\end{equation}
The central charge $c$ can be extracted as the coefficient of the quartic pole in the \eqref{TT_OPE}:
\begin{equation}
    c = \text{Vol}(D(\Omega))\frac{k(\Omega) \kappa^{\msf{a}\msf{a}}}{2k(\Omega)+C_2} = \text{Vol}(D(\Omega)) \frac{2k(\Omega)\text{dim}(\mathfrak{g})}{2k(\Omega)+2h^{\vee}} = \text{Vol}(D(\Omega)) \frac{k(\Omega)\text{dim}(\mathfrak{g})}{k(\Omega)+h^{\vee}}\,,
\end{equation}
in which we used
\begin{equation}
    \kappa^{\msf{a}\msf{b}}=2\delta^{\msf{a}\msf{b}}, \quad C_2 = 2h^{\vee} \,.
\end{equation}
The presence of a total number of $\text{Vol}(D(\Omega))$ symplectic bosons in the theory results in the divergence in the central charge.   

The canonical stress-energy tensor $T$ generates holomorphic translation $\partial_z$ on the sphere. Here we compute the OPE between the stress tensor $T$ and current algebra generators. 
\begin{align}
    &T(z_1)J^{\msf{a}}_{\Delta}(z_2)\nonumber\\
    &\sim \frac{1}{2k(\Omega)+C_2}\int\d^2q\,\d^2m\sum_{\msf{b}} \frac{k^{\msf{bd}}\delta_{-m+\Delta-q,0}}{(z_1-z_2)^2} \,\tilde\eta^{\msf{b}}_m\eta^{\msf{c}}_q\,f^{\msf{a}}{}_{\msf{cd}}+\phi\tilde\phi\,\text{terms}+\frac{1}{z_{12}^2}(-\Delta\tilde \phi^{\msf{a}}_{\Delta}(z_1))\nonumber
    \\&\sim\frac{1}{z_{12}^2}\,J^{\msf{a}}_{\Delta}(z_1)\sim\frac{1}{z_{12}^2}\,J^{\msf{a}}_{\Delta}(z_2)+\frac{1}{z_{12}}\partial_{z_2} J^{\msf{a}}_{\Delta}(z_2)\,,
\end{align}
where we have dropped the skew-symmetric simple pole terms and expanded around $z_1=z_2$ to include the subleading term. Similarly, one can compute the OPE between the stress tensor and $\tilde J^{\msf{a}}$:
\begin{align}
    T(z_1)\tilde J^{\msf{a}}_{\Delta}(z_2)
    \sim& \frac{1}{2k(\Omega)+C_2}\int\d^2q\,\d^2m\sum_{\msf{b}} \frac{k^{\msf{bd}}\delta_{-m+\Delta-q,0}}{(z_1-z_2)^2} \,\tilde\eta^{\msf{b}}_m\phi^{\msf{c}}_q\,f^{\msf{a}}{}_{\msf{cd}}\nonumber
    \\&+\frac{k^{\msf{bc}}\delta_{-m+q,0}}{(z_1-z_2)^2} \,\tilde\phi^{\msf{b}}_m\tilde\eta^{\msf{d}}_{\Delta-q}\,f^{\msf{a}}{}_{\msf{cd}}+\frac{1}{z_{12}^2}(-\Delta+2\im)\tilde \eta^{\msf{a}}_{\Delta}(z_1))\nonumber
    \\\sim&\frac{1}{z_{12}^2}\,\tilde J^{\msf{a}}_{\Delta}(z_1)\sim\frac{1}{z_{12}^2}\,\tilde J^{\msf{a}}_{\Delta}(z_2)+\frac{1}{z_{12}}\partial_{z_2} \tilde J^{\msf{a}}_{\Delta}(z_2)\,,
\end{align}
which indeed generates holomorphic translation as well. 

Note that conformal weight in the theory on $\mathbb{CP}^1_{[\mu\bar\lambda]=0}$ is not related to rescaling weight, or Mellin mode number $\Delta$. In the theory on $\mathbb{CP}^1_{[\mu\bar\lambda]=0}$, the currents $J, \tilde J$ are conformal primaries of conformal weight $1$, independent of their scaling weights, which is twice the real part of $-\Delta$. Similarly, the basic fields are conformal primaries of weight $0$ (for $\eta, \phi$) and $1$ (for $\tilde \eta, \tilde \phi$). This is the result of \eqref{a_restriction}, where we have completely eliminated the anti-holomorphic scaling weights. The benefit of this is that we end up with holomorphic 2d fields, which in the meantime completely decouples the conformal and scaling weights like in usual 2d CFTs. We could also kept the anti-holomorphic scaling weights and done the OPEs carefully tracking both scaling weights of all the fields. A more natural definition of the stress tensor in that case would involve a twist of the current form of $T$:
\begin{equation}
    T'= \frac{1}{2k(\Omega)+C_2}\int\d^2q\sum_{\msf{a}}:\tilde\eta^{\msf{a}}_{q}\,\partial\eta^{\msf{a}}_{(-q)}+\phi^{\msf{a}}_{q}\,\partial\tilde\phi^{\msf{a}}_{(-q)}-2\text{Re}(q)\,\partial\left(\tilde\eta^{\msf{a}}_{q}\,\eta^{\msf{a}}_{(-q)}+\tilde\phi^{\msf{a}}_{q}\,\phi^{\msf{a}}_{(-q)}\right):\,.
\end{equation}
where the twist is chosen in order to match the scaling weight with the conformal weights of fundamental fields. Under the twist, $\eta_{\Delta}, \phi_{\Delta}$ have conformal weight $\Delta$ while $\tilde \eta_{-\Delta}, \tilde \phi_{-\Delta}$ have conformal weight $1-\Delta$. It is not clear which prescription is more natural.

To further substantiate the 2d current algebra CFT \eqref{P1_current_action} as capturing 4d information, one needs to compute correlation functions purely based on the 2d CFT. Given the QSDYM we started with is free of 1-loop gauge anomaly, the only existing observable is the $++-$ three point function. This can be matched trivially by substituting in on-shell wavefunctions into the 3-point interaction. In order to enrich the theory with higher multiplicity observables, we shall insert marginal local gauge invariant operators which provides non-trivial background for the 2d correlator. We shall see that the form factors for MHV scattering can in fact be recovered from such insertions, in the language of 2D CFT correlators rather than 4D scattering amplitudes.


\section{Form factor/Correlation function match}\label{sec_ff_cf}
In this section we will focus on correlators of operator insertions that are on the locus $\mathbb{CP}^1_{[\mu \bar \lambda]=0}$ and that have explicit dependence on $\tilde \eta, \tilde \phi$. These are the correlators that are not accessible from a conventional treatment of QSDYM or full Yang-Mills on twistor space. In all the calculations that follow, the part of the action on $\mathbb{MT}$ decouples entirely because the operators we consider do not live on $\mathbb{MT}$, and only couple to things that do not live on $\mathbb{MT}$. In order to highlight this, we will put a subscript `current' on the correlators. We want to highlight that all calculations in this section can be done as 2D CFT calculations using \eqref{P1_current_action} on $\mathbb{CP}^1$ where we forget entirely about the theory on $\mathbb{MT}$. 

As discussed, we add the following gauge invariant operator (suppressing the sum over modes):
\begin{equation}\label{MHV_operator}
    S_{\text{MHV}}[b',a']:=\int_{\mathbb{CP}^1_{[\mu\bar\lambda]=0}} \D \lambda \wedge \text{tr}\left(b'_{\Delta}(\lambda_1) \,\frac{1}{\bar\D} \,b'_{-\Delta}(\lambda_2) \right)\la \lambda_1 \lambda_2 \ra^{3}\,.
\end{equation}
$S_{\text{MHV}}$ is motivated by a scaling reduction of the non-local MHV generating functional on $\mathbb{PT}^{\cO}$ in Euclidean signature, for which we refer the readers to appendix \ref{sec:appendix_A}. Since the discussion there was purely Euclidean, we do not treat it as a derivation of \eqref{MHV_operator}, instead just as a motivation. $S_{\text{MHV}}$ generates a deformation of the $\tilde J \tilde J$ correlator. The way to see this is to consider the formal substitution:
\begin{multline}
    \left\la \e^{S_{\text{MHV}}} \tilde J^{\msf{a}}_{\Delta_1}(z_1, \bar z_1)\tilde J^{\msf{b}}_{\Delta_2}(z_2,\bar z_2)  \right\ra_{\text{current}}\\
    =\left\la \e^{S_{\text{MHV}}} \frac{\delta}{\delta b^{'\msf{a}}_{-\Delta_1}(z_1,\bar z_1)}\frac{\delta}{\delta b^{'\msf{b}}_{-\Delta_2}(z_2,\bar z_2)}  \right\ra_{\text{current}}\,.
\end{multline}
In which the functional derivatives act on the exponentiated current algebra action \eqref{P1_current_action}. Note that they are only defined on the locus $[\mu \bar \lambda]=0$ and do not act on any non-localising $b$ living on $\mathbb{MT}$. One may therefore integrate by parts and act only on $\e^{S_{\text{MHV}}}$. We therefore have:
\begin{equation}
    \left\la \tilde J^{\msf{a}}_{\Delta_1}(z_1, \bar z_1)\tilde J^{\msf{b}}_{\Delta_2}(z_2,\bar z_2) \right\ra_{\text{current+MHV}}= \delta_{\Delta_1+\Delta_2,0}\left(\frac{1}{\bar\D}\right)^{\msf{ab}} (z_1-z_2)^{3}\,,
\end{equation}
where we have used local coordinates on the sphere. The weight matching delta function comes from the matching in scaling weights in $S_{\text{MHV}}$ and $\left(\frac{1}{\bar\D}\right)^{\msf{ab}}$ formally denotes the holomorphic inverse propagator on the sphere. Working in Woodhouse gauge to trivialise the connection on the sphere:
\begin{align}
    \left\la \tilde J^{\msf{a}}_{\Delta_1}(z_1, \bar z_1)\tilde J^{\msf{b}}_{\Delta_2}(z_2,\bar z_2)\right\ra_{\text{current+MHV}}&= \frac{\delta_{\Delta_1+\Delta_2,0}\,\kappa^{\msf{ab}}\d z_2}{z_1-z_2} (z_1-z_2)^{3} \nonumber\\
    &= \delta_{\Delta_1+\Delta_2,0}\,\kappa^{\msf{ab}}\d z_2 (z_1-z_2)^{2}\,. \label{2pt_tjtj}
\end{align}
We have assumed that the vev of $b$ vanishes, and therefore we can disregard the term where each functional derivative acts on the exponential, giving a $\la (\frac{1}{\bar D} b)(\frac{1}{\bar D} b)\ra$ type correlator (these contributions will play a role when there are more than 2 $\tilde J$ insertions, i.e at NMHV level and we must be careful). We see that in the presence of $S_{\text{MHV}}$ \eqref{MHV_operator}, a module is created in the current algebra system for the regular two point function $\tilde J\tilde J$. Also note that although the two point function $\tilde J\tilde J$ can be computed given the classical background \eqref{MHV_operator}, their OPE is still vanishing since there are no OPE between their component fields in the CFT Lagrangian \eqref{P1_current_action}. This reflects the intrinsic chiral nature of twistor theory. For example, even the N$^k$MHV tree-level amplitudes in $\mathcal{N}=4$ super Yang-Mills \cite{Witten:2003nn,Roiban:2004yf} derived from the twistor string does not have manifestly correct $--$ collinear splitting function until some ambidextrous integral transform is performed. Given the non-trivial two point function \eqref{2pt_tjtj}, we find that we can compute interesting non-vanishing current correlators. Consider the three current correlator of MHV type $\left\la\tilde J\tilde J J \right\ra$:
\begin{align}
    &\left\la \tilde J^{\msf{a_1}}_{\Delta_1}(z_1, \bar z_1)\tilde J^{\msf{a_2}}_{\Delta_2}(z_2,\bar z_2) J^{\msf{a_3}}_{\Delta_3}(z_3,\bar z_3) \right\ra_{\text{current+MHV}}\nonumber
    \\=&\,\frac{f^{\msf{a_1a_2a_3}}\delta_{\Delta_1+\Delta_2+\Delta_3,0}(z_1-z_2)^{2}}{(z_2-z_3)}-\frac{f^{\msf{a_1a_2a_3}}\delta_{\Delta_1+\Delta_2+\Delta_3,0}(z_1-z_2)^{2}}{(z_1-z_3)}\nonumber
    \\=&\frac{f^{\msf{a_1a_2a_3}}\delta_{\Delta_1+\Delta_2+\Delta_3,0}(z_1-z_2)^{3}}{(z_2-z_3)(z_1-z_3)}\,.
\end{align}
Using an inductive argument, the 2d correlator with two $\tilde J$ and $n-2$ insertions of $J$:
\begin{multline}
    \left\la \tilde J^{\msf{a_1}}_{\Delta_1}(z_1, \bar z_1)\tilde J^{\msf{a_2}}_{\Delta_2}(z_2,\bar z_2) \prod_{i=3}^n J^{\msf{a_i}}_{\Delta_i}(z_i,\bar z_i) \right\ra_{\text{current+MHV}}=\delta_{\boldsymbol{\Delta},0}\times
    \\ \sum_{\sigma \in S_{n}}\tr\left(T^{\msf{a}_{\sigma(1)}}T^{\msf{a}_{\sigma(2)}}T^{\msf{a}_{\sigma(3)}}\dots T^{\msf{a}_{\sigma(n)}}\right) \,\frac{(z_{\sigma(1)}-z_{\sigma(2)})^4}{(z_{\sigma(1)}-z_{\sigma(2)})(z_{\sigma(2)}-z_{\sigma(3)})\cdots(z_{\sigma(n)}-z_{\sigma(1)})}\,,
\end{multline}
where we recognize the last part as the familiar Parke-Taylor factor \cite{Parke:1986gb} and the sum runs over permutations $\sigma$. It is important to note that none of the pairs of $J^{\msf{a}}$ currents in the correlator have their weights sum to $0$, otherwise the infinite level term \eqref{JJ_OPE} appears\footnote{$\Delta_i+\Delta_j\neq 0$ for any $i,j\in\{3,4,\dots,n\}$.}. Notice that the correlator is subject to the constraint that the sum of all the scaling weights $\boldsymbol{\Delta}=\Delta_1+\Delta_2+\sum_i \Delta_i=0$, the current correlators can be seen to reproduce the $\tr B^2$ form factors for the MHV amplitudes in Yang-Mills in 4d Minkowski space. We computed the correlator purely based on a 2d CFT with marginal deformations living on the celestial sphere itself, which have no knowledge about the momentum conserving delta function. In order for it to appear, the 2d CFT needs to have information about the 4d bulk by extending the theory off of $\mathbb{CP}^1_{[\mu\bar\lambda]=0}$ into its formal neighborhoods. We have a more thorough discussion on this in the discussion section.

Associativity of the S-algebra can be compromised by 1-loop three point off-shell splitting functions \cite{Costello:2022upu}, which can be rectified by adding the axion on spacetime. We integrated out the axion in the counter term for the gauge anomaly in the upstairs twistor action \eqref{QSDYM_6d}, generating a counter term defined on $\mathbb{CP}^1 \times \mathbb{CP}^1$ in the scale reduced theory. 
\begin{equation}
    \int_{\mathbb{CP}^1_{[\mu\bar\lambda]=0}} \tr\left(\tilde \eta \,\partial_n a \phi \,\partial_n \phi +  \tilde \eta \,\partial_n a\,\frac{1}{\bar \partial}\, \partial_n a \,\partial_n \phi+\text{other terms} \right)\,,
\end{equation}
where we have incorporated the top holomorphic form $\D\lambda$ in $\tilde\eta$ and $\partial_n$ was a holomorphic partial differential pointing in the $\mathbb{C}^*$ reduction direction on $\mathbb{PT}^{\cO}$, becoming a mass term on $\mathbb{CP}^1_{[\mu \bar \lambda]=0}$. It is clear that this term will contribute to the current $J$, as it couples to $a$. More study is needed to see if our approach can reproduce the known results. 

\subsection{Higher degree form factors}
The 2d deformation operator \eqref{MHV_operator} provided non-trivial $\tilde J\tilde J$ two point function and higher multiplicity correlators of MHV type. In a similar fashion, one could insert the following deformation operator:
\begin{equation}\label{NMHV_operator}
    S_{\text{N$^{k-2}$MHV}}= \int_{\mathbb{CP}^1_{[\mu\bar\lambda]=0}}\D\lambda\wedge\tr\left(b'_{\Delta_1}\frac{1}{\bar\D}b'_{\Delta_2}\cdots\frac{1}{\bar\D}b'_{\Delta_k}\right)\la 12\ra^2\la 23\ra^2\cdots\la (k-1)k\ra^2\la k1\ra\,,
\end{equation}
where the sum of all the mode numbers $\sum_{i=1}^{k}\Delta_i=0$. Note that this is also required in order for the entire expression to make sense to integrate over $\mathbb{CP}^1$. The claim is that \eqref{NMHV_operator} provides non-zero value for correlation function of $k$ $\tilde J$ currents. Trading each $\tilde J$ insertion for a $b'$ functional derivative using the coupling, the $\tilde J$ correlator can be computed (suppressing color and mode numbers)
\begin{equation}\label{NMHV_correlator}
    \left\la\tilde J(z_1)\dots\tilde J(z_k)\right\ra_{\text{N$^{k-2}$MHV+current}} = \la 12\ra\la 23\ra\cdots\la (k-1)k\ra\la k1\ra\,.
\end{equation}
When $k=2$, we see that we recover the two point function \eqref{2pt_tjtj}. After inserting arbitrarily many $J$s, one could use the $J\tilde J$ OPEs to reduce the correlator to \eqref{NMHV_correlator}, which give the all-multiplicity N$^{k-2}$MHV form factors. It would be interesting to understand the connection with \eqref{MHV_operator} under a purely 2d OPE rule, which will be a 2d version of the celebrated CSW rule in Yang-Mills \cite{Cachazo:2004kj}.

An intriguing hint towards this goal is the following observation. Instead of the non-local $\bar\D^{-1}$ operators, one can consider adding an auxiliary pair of symplectic bosons $\rho_{m+\im \rho}^{\alpha\beta\delta}$, $\tilde\rho_{m+\im \rho}^{\alpha\beta\delta}\in\Omega^0(\mathbb{CP}^1,\cO(-2m-1)\times \mathfrak{g})$ to write a generating functional for the deformation operators (with the mode numbers suppressed as usual):
\begin{multline}
    S_{\rho\tilde\rho}:=\int_{\mathbb{CP}^1_{[\mu\bar\lambda]=0}}\D\lambda\wedge\text{tr}\left(\rho^{\alpha\gamma\xi}\left((\bar\partial+a')\epsilon_{\alpha\beta}\epsilon_{\gamma\delta}\delta^{\sigma}{}_{\xi}+b'\lambda_{\alpha}\lambda_{\beta}\lambda_{\gamma}\lambda_{\delta}\delta^{\sigma}{}_{\xi}\right)\tilde\rho^{\beta\delta}{}_{\sigma}\right.\\
    \left.\psi^{\alpha\gamma\xi}\left((\bar\partial+a')\epsilon_{\alpha\beta}\epsilon_{\gamma\delta}\delta^{\sigma}{}_{\xi}+b'\lambda_{\alpha}\lambda_{\beta}\lambda_{\gamma}\lambda_{\delta}\delta^{\sigma}{}_{\xi}\right)\tilde\psi^{\beta\delta}{}_{\sigma}+b'\left(\rho^{\alpha\beta\gamma}+\tilde\rho^{\alpha\beta\gamma}\right)\lambda_{\alpha}\lambda_{\beta}\lambda_{\gamma}\right)\,,
\end{multline}
where integrating out $\rho,\tilde\rho,\psi$ and $\tilde\psi$ manifestly gives a perturbative series containing the N$^{k-2}$MHV deformation operator \eqref{NMHV_operator} for each $k \geq 2$. Explicitly, we do Gaussian integral over the bosonic $\rho^{\alpha\beta\gamma}$ system and replace
\begin{equation}
    \tilde\rho^{\alpha\beta\gamma} \rightarrow \left(\left(\bar\partial+a'\right)\veps_{\alpha\sigma}\veps_{\beta\xi}\delta^{\delta}_{\gamma}+b'\lambda_{\alpha}\lambda_{\beta}\lambda_{\sigma}\lambda_{\xi}\delta^{\delta}_{\gamma} \right)^{-1}\, \lambda^{\alpha}\lambda^{\beta}\lambda^{\gamma}\,b' \,,
\end{equation}
where the inverse operator is a formal series:
\begin{align}
&\left(\left(\bar\partial+a'\right)\veps_{\alpha\sigma}\veps_{\beta\xi}\delta^{\delta}_{\gamma}+b'\lambda_{\alpha}\lambda_{\beta}\lambda_{\sigma}\lambda_{\xi}\delta^{\delta}_{\gamma} \right)^{-1} \\
= &\frac{\veps_{\alpha\sigma}\veps_{\beta\xi}\delta^{\delta}_{\gamma}}{\bar \partial + a} - \frac{1}{\bar \partial + a'} b'\lambda_{\alpha}\lambda_{\beta}\lambda_{\sigma}\lambda_{\xi}\delta^{\delta}_{\gamma}\frac{1}{\bar \partial +  a'} + ...
\nonumber
\\ = &\sum_{i=1}^{\infty}(-1)^{i+1} \frac{(\lambda_2)_{\alpha}(\lambda_2)_{\beta}}{\bar\partial + a'} b'(\lambda_2)\frac{\la 2 3 \ra^2}{\bar\partial + a'}b'(\lambda_3) \ldots \frac{\la i i+1 \ra^2}{\bar\partial + a'}b'(\lambda_{i+1})  \frac{(\lambda_{i+1})_{\sigma}(\lambda_{i+1})_{\xi}}{\bar \partial + a'} \delta^\delta_\gamma \nonumber \,.
\end{align}
We see that the inverse propagators $\left(\bar\D\,\veps_{\alpha_i\sigma_{i+1}}\veps_{\beta_i\xi_{i+1}}\delta^{\delta_i}_{\gamma_{i+1}}\right)^{-1}$ links indices of neighboring $\lambda$s and give $\la i(i+1)\ra^2$ except for $\lambda_1$ and $\lambda_n$, where two out of three have been contracted with $\lambda_2$ and $\lambda_{n-1}$ respectively, the remaining one pair up and give $\la n1\ra$, recovering the factors in \eqref{NMHV_operator}. The Gaussian integral over the $\psi, \tilde \psi$ system has no contribution except to cancel the inverse determinant $\text{det}\left(\left(\bar\partial+a'\right)\veps_{\alpha\sigma}\veps_{\beta\xi}\delta^{\delta}_{\gamma}+b'\lambda_{\alpha}\lambda_{\beta}\lambda_{\sigma}\lambda_{\xi}\delta^{\delta}_{\gamma} \right)$ that arises from the Gaussian integration over $\tilde \rho, \rho$. For N$^k$MHV order deformation operator, we simply pick the term with $k+2$ $b'$s from the expansion of the effective holomorphic propagator.

From another point of view, if we do not integrate out the auxillary fields, $S_{\rho\tilde\rho}$ alters the $\tilde J$ correlator by altering the $b'$ functional derivatives such that they now pull down insertions of 
\begin{align}
\la\rho\lambda\lambda,\tilde\rho\lambda\lambda\ra& + \la\psi\lambda\lambda,\tilde\psi\lambda\lambda\ra + (\tilde\rho+\rho,\lambda\lambda\lambda) := 
\\ &\rho^{\alpha\gamma\xi}\lambda_{\alpha}\lambda_{\beta}\lambda_{\gamma}\lambda_{\delta}\delta^{\sigma}{}_{\xi}\tilde\rho^{\beta\delta}{}_{\sigma} + \psi^{\alpha\gamma\xi}\lambda_{\alpha}\lambda_{\beta}\lambda_{\gamma}\lambda_{\delta}\delta^{\sigma}{}_{\xi}\tilde\psi^{\beta\delta}{}_{\sigma} + \left(\rho^{\alpha\beta\gamma}+\tilde\rho^{\alpha\beta\gamma}\right)\lambda_{\alpha}\lambda_{\beta}\lambda_{\gamma}\,, \nonumber
\end{align}
and these insertions in the correlator can have OPEs with each other via the $\rho \tilde\rho$ OPE and the $\psi \tilde\psi$ OPE. Suppressing colors and mode number:
\begin{multline}
    (\la\rho\lambda\lambda,\tilde\rho\lambda\lambda\ra + \la\psi\lambda\lambda,\tilde\psi\lambda\lambda\ra+ (\tilde\rho+\rho,\lambda\lambda\lambda))_{\Delta_1} (\la\rho\lambda\lambda,\tilde\rho\lambda\lambda\ra + \la\psi\lambda\lambda,\tilde\psi\lambda\lambda\ra+ (\tilde\rho+\rho,\lambda\lambda\lambda))_{\Delta_2} 
    \\ \sim 2\la 1 2 \ra^2 \delta_{\Delta_1+\Delta_2,0} + \la 12 \ra \left(\la\rho(\lambda_1)\lambda_1\lambda_1,\tilde\rho(\lambda_2)\lambda_2\lambda_2\ra + \rho(\lambda_1)\lambda_2\lambda_1^2 + \ldots \right) + \text{terms with $b'$}\,.
\end{multline}
The double zero term has a finite coefficient despite the infinite contributions from the double contractions of the quadratic terms in $ \rho, \tilde \rho$ and $\psi, \tilde \psi$ because they are equal and opposite (this is best seen from taking an $\Omega$-truncation and letting $\Omega \rightarrow \infty$). The only remaining contribution is the finite contribution of the single contraction between terms linear in $\rho, \tilde \rho$. This double zero term reproduces the $\la \tilde J \tilde J\ra$ correlator, as expected. The simple zero terms that are linear in $\rho, \tilde \rho$ contribute to the higher multiplicity $\tilde J$ correlator, e.g if there was a $\tilde \rho(\lambda_3) \lambda_3^3$ from another insertion of $\tilde J$ we would get 
\begin{equation}
    \left\la \la 12 \ra\rho(\lambda_1)\lambda_2\lambda_1^2 \left(\tilde\rho(\lambda_3)\lambda^3_3\right) \right\ra= \left\la \la 12 \ra\frac{\la23\ra\la 1 3\ra \la 1 3\ra}{\la 1 3\ra} + \mathcal{O}(b')\right\ra = \la 12 \ra\la23\ra\la 1 3\ra\,,
\end{equation}
where we have used the OPE between $\rho$ and $\tilde\rho$, which has a simple pole. Further insertions of $J$s in the correlator would give NMHV form factors. This can be done for all N$^k$MHV configurations. 

$S_{\rho \tilde\rho}$ provides a local expression for the deformation operator, at the expense of introducing an additional system in the path integral. It would be interesting to explore this further, instead of integrating out the pair of bosons, we can also integrate out the modes $a$ and $b$ themselves, which in the supersymmetric case produced some matrix model generating various determinant operators found in giant graviton expansions \cite{Caron-Huot:2023wdh}.

\section{Discussion/Speculations}\label{sec:discussion}
In this paper, we formally split SDYM on $\mathbb{PT}^{\cO}$ through `large gauge transformations', where part of the action localises to the locus $[\mu\bar\lambda]=0$. After a $\mathbb{C}^*$ scaling reduction and retaining all the modes, the localised action provides an explicit realization of the infinite dimensional S-algebras. We examined the 2d correlators computed in the vacuum of an additional marginal operator, which captured the MHV form factor of 4d Yang-Mills theory.

With the 2d Lagrangian description of a CFT realising the S-algebra, we were able to compute its stress tensor and central charge. The central charge was found to be proportional to the size of the spectrum of the CFT. Given the scaling reduction was performed along a non-compact direction, the CFT is actually irrational, which renders the central charge to be $\infty$. This precisely reflects the fact that there are no dimensionless scales present in flat space. This has been observed in other constructions as well \cite{Stieberger:2023fju,Costello:2023hmi}. In principle, our procedure would streamline for integrable conformal theories in Minkowski space, preferably with known twistor space uplifts. Fully solvable codimension 2 models have been proposed in the literature \cite{Kapec:2022xjw,Duary:2022onm}, where a 2d/0d dictionary was established. Compared to this, our construction allows a 2d CFT for manipulation, where interesting dynamical phenomenon could arise.

\medskip

It is natural to ask whether this derivation of the 2d CFT could be recognized as a step $0$ progress towards a construction of the celestial CFT, as a weak-weak duality, where observables match after deformation of marginal operators. We shall point out the ambiguities present in the construction and important physical questions which require further clarifications. 

\medskip

The theory after $\mathbb{C}^*$ reduction can be seen as living on the minitwistor space of EAdS$_3$/LdS$_3$ with the $\mathbb{CP}^1$ at $[\mu\bar\lambda]=0$\footnote{It can be mapped to the celestial sphere where $x^2=0$ cuts $\scri$.}, which is most conveniently seen in embedding space metric. Although we have captured the Milne and Rindler region with $x^2>0$ and $x^2<0$ patches as discussed in appendix \ref{appendix:detail_scaling_reduction}, and $x^2=0$ with the celestial sphere at $u=0$ on $\scri^+$ or $v=0$ on $\scri^-$. The division between Milne and Rindler region is fixed by the light-cone of an arbitrarily chosen point $p$ in Minkowski space hence should not affect the physical observables. Although we have made explicit choice of $p$ to be the origin, given the fact that the theory we are considering is conformal, we could always use 4d conformal transformations to map other choices of $p$ back to the origin. To completely reconstruct Minkowski space, matching conditions for observables on different patches along the common boundary $x^2=0$ will have to be made clear. Observables constructed in different regions should be glued back together to give 4d scattering amplitudes, agnostic about the choice of $p$.

The other subtlety involves the Rindler region in general. It is worth noting that although we were able to perform scaling reduction on the patch $x^2<0$, which gives an action functional formally living on a minitwistor space for the LdS$_3$ slices \cite{LeBrun:2008ch} together with a 2d CFT living on the boundary celestial sphere $\mathbb{P}^1_{[\mu\bar\lambda]=0}$, the geometric importance of $[\mu\bar\lambda]=0$ is far from understood in this case compared to that in Euclidean AdS$_3$. As discussed in \cite{Bu:2023cef}, $[\mu\bar\lambda]=0$ degenerates the geodesics in AdS$_3$ and one lands on the boundary bitwistor space $A_{\mathbb{E}}\cong\mathbb{CP}^1$. In LdS$_3$ however, the geometrical relation between its boundaries and the corresponding $A_{\mathbb{E}}$ is not known. Besides the geometry, the generic lack of understanding of the correspondence between $\mathbb{MT}$ and dS$_3$ is preventing us from making explicit statements about the role of 4d large gauge parameter or the gluon fields themselves on dS$_3$ slices. To our knowledge, the (linearised) physical degrees of freedom in dS$_3$ have not been explicitly constructed as cohomology classes on $\mathbb{MT}$. Hence a first step towards a complete understanding of the gluing and matching 4d observables in different regions of $\mathbb{M}$ requires our full understanding of the de Sitter slice and its minitwistor space.

We have picked the gauge profile on $\mathbb{PT}^{\cO}$ with $\myd=1$ \eqref{PT_1_gold} to derive the current algebra CFT. It is worth noting that this is by no means the most generic form of a `large gauge parameter' on $\mathbb{PT}^{\cO}$ even with arbitrary $\myd$ \eqref{eq2}. In principle, any almost $\bar\partial$ exact (allowing exceptions at isolated points) profile on $\mathbb{PT}^{\cO}$ with homogeneity $0$ in $\mu$ is permitted as they Sparling transform to pure gauge connections of order $\cO(r^0)$. A more complete analysis can be found in appendix \ref{subsappend:geometry}. In particular, as mentioned in section \ref{sec:large_gauge_4d}, higher integer $\myd>1$ gauge parameters localise the action to $(\myd-1)$th order formal neighborhoods of $\mathbb{CP}^1_{[\mu\bar\lambda]=0}$, which allows the current algebra CFT to have knowledge about the bulk. We were able to compute form factors in absence of momentum conserving delta functions from 2d correlation functions, the extension into the bulk through formal neighborhoods could potentially lead to the recovery of the missing delta functions, hence giving a more complete reconstruction of 4d dynamics.  

Finally, one wonders if similar procedure would work on Minkowski space itself rather than on its twistor space. The advantage of being on twistor space is the access to additional gauge redundancies, which conveniently allows us to pick special singular gauge parameters localising to the locus of desired projective lines. Other than this, ambiguities generally occur under large gauge transformations. It is questionable whether the theory after large gauge transformation is still equivalent to the original on the quantum level. This is indeed potentially an issue on spacetime, given the asymptotic scattering states could require additional boundary conditions since the additional gauge parameters do not fall off at infinity. On twistor space however, the gauge transformations are not `large' themselves, the $\mathbb{CP}^1$ at $[\mu\bar\lambda]=0$ we have localised the action to is not on the boundary of twistor space, but can be mapped to the celestial sphere on $\scri$ on spacetime. Hence there is no ambiguity with the definition of on-shell scattering degree of freedom on twistor space (the $H^{0,1}$ cohomology classes). On spacetime, if we were to follow a similar recipe, start by taking a special large gauge transformation (a Heaviside step function perhaps) which localises to the celestial sphere. After appropriate field redefinitions, a further scaling reduction gives a copy of the theory living on a 3d slice with another piece living on the celestial sphere. Although it is not completely clear how various BMS symmetries and charge conservation would function in this scenario, it is nevertheless conceivable that our recipe on twistor space could be realised on Minkowski space itself after addressing these potential issues.

\medskip

We further comment on the geometric relations of (mini-)twistor spaces appeared in this paper and those in \cite{Costello:2022wso}, where the authors reduced SDYM \eqref{SDYM_6d} action on the Euclidean real slice of twistor space: $\mathbb{PT}^{\cO}\cong\mathbb{R}^4\setminus\{0\}\times\mathbb{CP}^1$ on an $S^3\subset\mathbb{R}^4\setminus\{0\}$. Although losing the physical intuition of the Lorentzian signature Minkowski space, KK reduction yielded the twisted 3d $\mathcal{N}=2$ theory \cite{Aganagic:2017tvx} on $\mathbb{CP}^1\times\mathbb{R}_+$. Note that the origin of $\mathbb{R}^4$ has been removed from the beginning. The minitwistor space $\mathbb{MT}$ of EAdS$_3$ and LdS$_3$ can be recognized as a projective quadric embedded in $\mathbb{PT}^{\cO}\setminus\{0\}$, since all quadric $Q\in\mathbb{CP}^3$ are biholomorphic to $\mathbb{CP}^1\times\mathbb{CP}^1$, $\mathbb{PT}^{\cO}/\mathbb{C}^{*}\cong\mathbb{CP}^1\times\mathbb{CP}^1\setminus\{\lambda_\alpha=0\,\text{or}\,\mu^{\dal}=0\}$. Actual minitwistor space requires an additional $\mathbb{CP}^1$ parametrized by $\mu^{\dal}\propto\bar\lambda^{\dal}$ to be taken out, which can be mapped to the celestial at $u=0$ on $\scri^+$ or equivalently $v=0$ on $\scri^-$. To relate to the Euclidean picture, one can begin with $\mathbb{R}^4\setminus\{0\}\times\mathbb{CP}^1\cong S^3\times\mathbb{R}_+\times\mathbb{CP}^1$ and recognize $S^3$ as a Hopf fibration over $S^2\cong\mathbb{CP}^1$. Focusing on $\mathbb{CP}^1\times\mathbb{CP}^1$ as the base manifold, the circle bundle $S^1$ together with the $\mathbb{R}_+$ factor combines into $\mathbb{R}^2\setminus\{0\}\cong\mathbb{C}^*$. This is the non-compact direction we performed the scaling reduction to minitwistor space, where the modes are labelled by Mellin parameter and a discrete Fourier basis $\Delta\in\mathbb{Z}+\im\mathbb{R}$. Further discussions on the geometry in various signatures can be found in the appendix \ref{appendix:detail_scaling_reduction}.

\medskip

We will also further discuss directions beyond this work, including perhaps some speculative comments.

\paragraph{Immediate follow ups}
As discussed in section \ref{sec_ff_cf} and appendix \ref{sec:appendix_A}, insertion of the gauge invariant local operator \eqref{MHV_operator} computing MHV form factor in $\mathbb{M}$ is inspired by the MHV generating functional on $\mathbb{PT}^{\cO}$. Given addition of the MHV vertex on $\mathbb{PT}^{\cO}$ gives entire tree-level Yang-Mills S-matrix \cite{Boels:2006ir,Boels:2007qn,Cachazo:2004kj,Adamo:2011cb} using CSW rule, it is reasonable to hope that the addition of local operators \eqref{NMHV_operator} to the 2d current algebra CFT could give all form factors in full Yang-Mills in $\mathbb{M}$. This amounts to developing 2d CFT analogue of the CSW rules for Yang-Mills N$^k$MHV amplitudes \cite{Cachazo:2004kj}. It would interesting to furnish it with a derivation from the deformation operator of lower MHV degree \eqref{MHV_operator}, with no reference to the bulk. It would also be interesting to further investigate the symplectic boson $\rho\tilde\rho$ system inserted to allow a local deformation operator.

Besides gauge theory, we can also aim to replicate the recipe for self-dual Einstein gravity, where the 2d CFT is expected to generate the loop algebra of $w_{1+\infty}$ algebra \cite{Strominger:2021mtt}. The 1-loop gauge anomaly has been rectified after addition of the same axion \cite{Bittleston:2022nfr}, we could follow the same recipe of large gauge transformation and scale reduce. However, the further addition of marginal operators become slightly ambiguous, as there is no notion of `local gauge invariant operators' in gravity\footnote{Or in the context of Minkowski space scattering amplitudes, there is no known CSW rules for building tree level N$^k$MHV amplitudes in Einstein gravity.}. \cite{Bittleston:2022jeq} has proposed to consider operators of higher ghost number in the BRST cohomology and \cite{Sharma:2021pkl} has provided a smeared anti-self-dual perturbation off of self-dual Einstein gravity in $\mathbb{M}$, which are worth further investigations in the 2d CFT context provided that we could derive it from SD gravity on $\mathbb{PT}^{\cO}$. 

Yet another interesting question arises in the gravity case, a stress tensor could be derived in the current algebra CFT for SD gravity given a Lagrangian description. It has been observed in graviton scattering amplitudes in $\mathbb{M}$ that shadow transform of the subleading conformally soft graviton acts on the rest of Mellin transformed scattering amplitudes as a 2d CFT stress tensor. It would be worthwhile to compare whether the stress tensor of the derived 2d CFT matches this existing observation from the 4d side. 

\paragraph{Relations to top-down constructions}
The recently proposed topological B-model with target space being the twistor space $F$ of Burns space \cite{Costello:2022jpg,Costello:2023hmi} has presented a complete top-down example of twisted holography in asymptotically flat spacetime. It would be interesting to ask whether our recipe for deriving the 2d CFT could be used in $F$. The notion of pure gauge large gauge transformations would still be valid on Burn space as it is asymptotically Euclidean. Although a careful examination is needed for the slicing structures of Burns space compared to Minkowski space and the geometric structure of minitwistor spaces and their embeddings in $F$. In particular, the choice of the form of the singular `large gauge' parameter is heavily motivated by the geometry of minitwistor space of EAdS$_3$. $F\cong SU(2)\cong \text{AdS}_3\times S^3$ is the space of oriented geodesics in $\mathbb{CP}^2$ with an additional divisor $\mathbb{CP}^1$ at $\infty$. The minitwistor spaces we have considered are spaces of oriented geodesics in EAdS$_3$, which are orbit spaces of $SU(2)$ under quotients. After addressing these issues, it would certainly be interesting to repeat the procedure in this paper on $F$, starting with the gauge theory sector, with the hope to match the large $N$ chiral algebras living on the D1-D5 brane system in the open string sector of the B-model. 

\paragraph{Relation to other work}
In their seminal work \cite{deBoer:2003vf}, de Boer and Solodukhin performed KK reduction on free scalar theory on $(d+2)$-dimensional Minkowski space. Allowing arbitrary KK modes on $(d+1)$-dimensional Euclidean AdS or Lorentzian dS slices amounts to consider the full theory on Minkowski space. They established Green's function in AdS$_{d+1}$ and dS$_{d+1}$, which matched the scalar propagator in Minkowski space by virtual of the embedding space formalism. With the known AdS/CFT dictionary, they mapped two AdS bulk to boundary propagators onto two point correlation function of CFT$_{d}$ on boundary $S^d$. Our stories are inspired by \cite{deBoer:2003vf}, which can be seen as a step one effort towards realising their construction from the level of the off-shell action functional. Although our construction only works for conformal theories in four dimensions which admit twistor constructions so far, it is conceivable that the spacetime analogue can be worked out and generalised to other dimensions.

SDYM is known to be related to various lower dimensional integrable systems through symmetry reductions \cite{Ablowitz1993}. Although the purpose there is to study the integrable systems themselves as zero modes of the reduction, our derivation is more in the spirit of \cite{Bittleston:2020hfv}, where they examined a web of integrable theories from the level of the action functional with framework developed in \cite{Costello:2017dso}.


\paragraph{AdS$_3$/CFT$_2$ duality}
It is interesting to interpret our result in section \ref{sec_ff_cf} as a bulk spin-1/boundary spin-0 gauge parameter duality. At MHV degree, the boundary current correlator for currents $\tilde J, J$ that couple to the boundary value of the bulk fields $b,a$ can be read as a very restricted subset of the family of bulk amplitudes that the authors calculated in \cite{Bu:2023cef}. It is possible that consideration of large gauge parameters involving higher powers $\myd\in\mathbb{Z}_+$ \eqref{eq2} allows us to find the currents that couple to other bulk modes (with generic weights) and therefore recover current correlators that reproduce all of the bulk amplitudes. This could have a cute interpretation as an AdS$_3$/CFT$_2$ weak-weak duality in which a parent 4d theory provides the unifying construction for the duality.

The analogous statement in gravity would relate a bulk spin-2 field with the boundary spin-1 gauge parameter that controls large boundary diffeomorphisms.

\acknowledgments
It is a pleasure to thank Tim Adamo, Roland Bittleston, Eduardo Casali, Temple He, Sonja Klisch, Lionel Mason, Atul Sharma, David Skinner and Bin Zhu for interesting discussions and Eduardo Casali and Atul Sharma for working on related subjects. We also thank the anonymous JHEP referee for valuable suggestions and questions in the referee report for \cite{Bu:2023cef}. WB is supported by the Royal Society Studentship. SS is supported by the Trinity Internal Graduate Studentship. The work of SS has been supported in part by STFC HEP Theory Consolidated grant ST/T000694/1.

\newpage
\appendix 
\section{Full Yang Mills}\label{sec:appendix_A}
It is known that the deformation to QSDYM that gives the twistor action for Yang-Mills on $\mathbb{R}^4\setminus\{x=0\}$ (in perturbation theory) is the following \cite{Boels:2006ir}:
\begin{align}
 &S_{\text{YM}}:=S_{\text{QSDYM}}+S_{BB}
 \\
 &S_{BB}:=\int_{x\neq 0} \d^4x\, \D \lambda\, \text{tr}\left(\kB(1) \frac{1}{\bar \D} \kB(2)\right) \la \lambda_1 \lambda_2 \ra^3   \,,
\end{align}
where $\kB$ and the inverse propagator are restricted to their components that point along the $\mathbb{CP}_{\lambda}^1$. The correction term $S_{BB}$ is gauge invariant. In order to descend to $\mathbb{MT}$, it helps to rewrite the measure:
\begin{equation}
    S_{BB}=\int_{\mathbb{PT}^{\cO}\setminus\{\lambda=0\, \text{or} \,\mu=0\}} [\d \mu \d \mu] \D \lambda  \frac{[\d\hat \mu \d \hat \mu]}{\la \lambda \hat \lambda \ra^2}\, \text{tr}\left(\kB(1) \frac{1}{\bar \D} \kB(2)\right) \la \lambda_1 \lambda_2 \ra^3   \,.
\end{equation}
As always \cite{Bu:2023cef}, we first take the $\mathbb{R}^+$ scaling reduction corresponding to the Euler vector field (keeping all the modes), and then the $U(1)$ circle reduction (keeping all the modes) corresponding to the $S^1$ fibre over the resulting $S^3$, for a total of a $\mathbb{C}^*$ reduction, where we collect the continuous real mode index and the discrete real mode index into a complex parameter $\Delta$:
\begin{equation}
    \{\mathbb{R}^4\setminus\{0\}\} \times \mathbb{CP}^1 \xrightarrow{\mathbb{R}^+ \text{ reduction}} S^3 \times \mathbb{CP}^1 \xrightarrow{S^1 \text{ reduction}} S^2 \times \mathbb{CP}^1 \cong \mathbb{CP}^1_{\mu} \times \mathbb{CP}^1_{\lambda}\,.
\end{equation}
The vector fields on $\mathbb{PT}^{\cO}$ corresponding to each of the scalings is:
\begin{equation}
\mathbb{R}^+ \text{ action: } \left[\mu \frac{\partial}{\partial \mu}\right] + \left[\hat \mu \frac{\partial}{\partial \hat \mu}\right], \quad S^1 \text{ action: } \im\left[\mu \frac{\partial}{\partial \mu}\right] - \im\left[\hat \mu \frac{\partial}{\partial \hat \mu}\right] \,.
\end{equation}
We decompose every field in the integrand into a sum over modes under each reduction and perform the integral with the 1-form dual to each of these fields. The measure therefore loses the 2 form dual to $ \left[\mu \frac{\partial}{\partial \mu}\right] \wedge \left[\hat \mu \frac{\partial}{\partial \hat \mu}\right]$. Therefore (suppressing the sum over the modes), we can write the reduction of $S_{BB}$ as (recalling that we call the component of $\kB$ that points along the $\D \hat \lambda$ direction little $b$):
\begin{equation}
    \int_{\mathbb{CP}^1 \times \mathbb{CP}^1} [\mu \d \mu] \D \lambda \frac{[\hat \mu \d \hat \mu]}{\la \lambda \hat \lambda \ra^2}\, \text{tr}\left(b(1) \frac{1}{\bar \D} b(2)\right) \la \lambda_1 \lambda_2 \ra^3   \,.
\end{equation}
We would prefer to write the reduction of $S_{BB}$ as a sum of its a contribution to $\mathbb{MT}$ and to $\mathbb{CP}^1_{[\mu \bar \lambda]=0}$:
\begin{equation}
   \int_{\mathbb{MT}} [\mu \d \mu] \D \lambda \frac{[\hat \mu \d \hat \mu]}{\la \lambda \hat \lambda \ra^2}\, \text{tr}\left(b(1) \frac{1}{\bar \D} b(2)\right) \la \lambda_1 \lambda_2 \ra^3 + \int_{\mathbb{CP}^1_{[\mu \bar \lambda]=0}} \D \lambda \, \text{tr}\left(b'(1) \frac{1}{\bar \D} b'(2)\right) \la \lambda_1 \lambda_2 \ra^3   \,.
\end{equation}
And now it is in a form that we can add to our split theory on $\mathbb{MT}$ and $\mathbb{CP}^1_{[\mu\bar\lambda]=0}$. Adding it in, we conjecture the CCFT for full Yang Mills to be:
\begin{multline}\label{conjecture}
    S_{\text{current+MHV}} = S_{\text{current}}+S_{\text{MHV}} \\ = \int_{\mathbb{CP}^1_{[\mu\bar\lambda]=0}}\tr\left(-\eta\bar\partial\tilde\eta+a' J+\tilde\phi\bar\partial\phi+b'\tilde J\right) + \int_{\mathbb{CP}^1_{[\mu \bar \lambda]=0}} \D \lambda \, \text{tr}\left(b'(1) \frac{1}{\bar \D} b'(2)\right) \la \lambda_1 \lambda_2 \ra^3   \,,
\end{multline}
where we call the deformation term $S_{\text{MHV}}$, which in detail is (suppressing the sum over modes):
\begin{equation}
    S_{\text{MHV}}[b',a']:=\int_{\mathbb{CP}^1_{[\mu\bar\lambda]=0}} \D \lambda \wedge \text{tr}\left(b'_{\Delta}(\lambda_1) \,\frac{1}{\bar\D} \,b'_{-\Delta}(\lambda_2) \right)\la \lambda_1 \lambda_2 \ra^{3}\,.
\end{equation}
It is interesting to observe that the form of it looks like $S_{BB}$, the original deformation to QSDYM, except with the integral over $\mathbb{R}^4$ removed. In $S_{BB}$'s role as the on-shell generating functional for MHV diagrams, the integral over $\mathbb{R}^4$ was what gave the momentum conserving delta function, while the remaining part of the integrand, $S_{\text{MHV}}$, gave the form factor.

\section{Details of the scaling reduction}\label{appendix:detail_scaling_reduction}
As promised, we spell out the details of $\mathbb{C}^*$ scaling reduction of \eqref{Split_action_PT} on $\mathbb{PT}^{\cO}$, which was used extensively in the paper.

\subsection{Geometry of $\mathbb{PT}^{\cO}$ under the scaling reduction}\label{subsappend:geometry}

We wish to identify points on $\mathbb{PT}^{\cO}$ related under the $\mathbb{C}^*\cong\mathbb{R}_+\times S^1$ action on $\mu^{\dot \alpha}$. The vector fields on $\mathbb{PT}^{\cO}$ corresponding to each of the scalings is:
\begin{equation}
\mathbb{R}^+ \text{ action: } \left[\mu \frac{\partial}{\partial \mu}\right] + \left[\hat \mu \frac{\partial}{\partial \hat \mu}\right], \quad S^1 \text{ action: } \im\left[\mu \frac{\partial}{\partial \mu}\right] - \im\left[\hat \mu \frac{\partial}{\partial \hat \mu}\right] \,.
\end{equation}
\paragraph{Euclidean interpretation}
The interpretation of the $\mathbb{R}^+$ rescaling is most clear in the description of $\mathbb{PT}^{\cO}$ as the projective spin bundle over real Euclidean space,  $\mathbb{R}^4\setminus\{0\}\times \mathbb{P}^1$, coordinatised by $(x^{\alpha \dot\alpha},\lambda_{\alpha})$, where $x^{\alpha \dot\alpha}$ is a real Euclidean point $x^{\alpha \dot\alpha}=\hat x^{\alpha \dot\alpha}$:
\begin{equation}
    x^{\alpha \dot\alpha}:=\frac{\hat \mu^{\dot\alpha} \lambda^{\alpha} - \mu^{\dot\alpha}\hat\lambda^{\alpha}}{\la \lambda\hat\lambda\ra}\,.
\end{equation}
With this physical understanding, it is clear that under the $\mathbb{R}^+$ we have $x^{\alpha \dot\alpha}$ identified up to real positive rescalings, and therefore representatives can be chosen such that the set of equivalence classes $x \sim r x$ is an $S^3$ whose choice of radius amounts to a gauge fixing. Without loss of generality, consider the gauge slice $x^2=1$. The $U(1)$ reduction is then seen to be opposite pure phase rescalings of the two terms in $x^{\alpha \dot\alpha}$, which amounts to motion along the great circle geodesic $S^1$. Therefore in this picture, $\mathbb{PT}^{\cO}/\mathbb{C}^*$ is recognised to be the space of geodesics on $S^3$.

\paragraph{Lorentzian interpretation} As a quick refresher, the subset of $\mathbb{PT}^{\cO}$ that corresponds to null geodesics on $\mathbb{M}$ is a 5 real dimensional subset known as $\mathbb{PN}$, which is the zero locus of the real constraint on the bilinear $Z \cdot \bar Z:= [\mu \bar\lambda] + \la \bar \mu \lambda \ra$:
\begin{equation}
    \mathbb{PN} := \{Z\cdot \bar Z = 0| Z^I \in \mathbb{PT}^{\cO}\}\,.
\end{equation}
The barred (Lorentzian) conjugation interchanges the dotted and undotted spin bundles:

\begin{equation}
    \bar \lambda^{\dot \alpha}:=\begin{pmatrix}
        \bar \lambda^{0} \\ \bar \lambda^1
    \end{pmatrix}, \bar \mu^{\alpha} :=\begin{pmatrix}
        \bar \mu^{\dot 0} \\ \bar \mu^{\dot 1}
    \end{pmatrix}\,.
\end{equation}
Via the incidence relations $\mu = \im x \lambda$ (the factor of $\im$ is in keeping with the convention and simplifies expressions later) for a real Lorentzian point $x^{\alpha \dot\alpha}=\bar{x}^{\alpha \dot\alpha}$, a point in $\mathbb{PN}$ corresponds to a a null geodesic on $\mathbb{M}$:
\begin{align}
    &\mu^{\dot\alpha}=\im x^{\alpha \dot\alpha} \lambda_{\alpha}, \quad x^{\alpha \dot\alpha}=\bar{x}^{\alpha \dot\alpha} \label{B.4}\,;
    \\
    &x^{\alpha \dot\alpha}(t) = x_0^{\alpha \dot\alpha} + t \lambda^{\alpha} \bar\lambda^{\dot\alpha} \subset \mathbb{M}, \quad t\in\mathbb{R}\label{B.5}\,,
\end{align}
which can be seen by contracting $\bar\lambda_{\dal}$ on both sides of \eqref{B.4} and observing that the real part is equal to $Z \cdot \bar Z$ and in fact vanishes. The incidence relation is preserved under real Lorentzian translations of $x$ in the $\lambda\bar\lambda$ direction, and therefore the incidence relation doesn't specify real Minkowski points but rather the real null geodesic \eqref{B.5}. 
Note that the $Z\cdot\bar Z = 0$ condition can be written as the condition that the following $w\in\mathbb{R}$:
\begin{align}
    &w := \frac{x^{\alpha \dot\alpha}(t)\lambda_{\alpha}\bar\lambda_{\dot\alpha}}{\la\lambda\hat\lambda\ra} = \frac{x^{\alpha \dot\alpha}(0)\lambda_{\alpha}\bar\lambda_{\dot\alpha}}{\la\lambda\hat\lambda\ra} = -\im\frac{[\mu \bar\lambda]}{\la\lambda\hat\lambda\ra} \in \mathbb{C}\,;
    \\
    &Z\cdot \bar Z = \la \lambda \hat\lambda\ra\left(\im w-\im\bar w\right)=2\im\la \lambda \hat\lambda\ra \text{Im}(w)\,.
\end{align}
The alternative understanding that most naturally gives us insight into the $\mathbb{M}$ picture is as follows. First consider the action of the $U(1)$ rescaling on the bilinear $Z \cdot\bar Z \propto \text{Im}(w)$. For points $(\mu^{\dot\alpha},\lambda_{\alpha})$ with $[\mu \bar\lambda]\neq0 \iff w \in \mathbb{C}^*$, the orbit under the $U(1)$ action contains two points on $\mathbb{PN}$, related by the $\mathbb{Z}_2$ that sends $w \rightarrow -w$. Therefore we have an $S^1$ fibration of $\{\mathbb{PT}^{\cO}\setminus \{[\mu \bar\lambda]=0\}\}$ over $\{\mathbb{PN}\setminus \{w=0\}\}/\{w\sim -w\}$, which is the space of pairs of null geodesics on $\mathbb{M}$ related by sending $w \rightarrow -w$.

Now we must do the $\mathbb{R}^+$ reduction. Consider the $\mathbb{R}^+$ orbit of a point in $\mathbb{PN}\setminus \{w=0\}$. Via the incidence relations, this corresponds to the $\mathbb{R}^+$ orbit of its null geodesic in $\mathbb{M}$, which is the open half plane that contains the null geodesic and limits to but does not contain $x \sim \lambda^{\alpha}\bar \lambda^{\dot \alpha}$. For the space $\{\mathbb{PN}\setminus \{w=0\}\}/\{w\sim -w\}$, the $\mathbb{R}^+$ orbit consists of the union of the half plane and the image of the half plane under $w \rightarrow -w$, which is the full plane that contains both null geodesics, with the line $x \sim \lambda^{\alpha}\bar \lambda^{\dot \alpha}$ removed. $\mathbb{MT}$ as a locus in $\mathbb{M}$ is therefore the space of 2-planes that contain two null geodesics related by $w\rightarrow -w$, with the line $x \sim \lambda^{\alpha}\bar \lambda^{\dot \alpha}$ removed. In the embedding space language, we see that such planes intersect $\mathbb{H}_3$ slices in the Milne region of $\mathbb{M}$ in spacelike geodesics (which are all the geodesics on $\mathbb{H}_3$), and intersect the dS$_3$ slices in the Rindler regions of $\mathbb{M}$ in timelike geodesics. This gives rise to the usual slogan that the minitwistor space is the space of geodesics in hyperbolic space.

Another instructive point of view to take is that $\mathbb{MT}$ as a locus in $\mathbb{M}$ is defined by the ordered pair of two null geodesics passing through the origin, and therefore is fixed by 2 points on the projective null cone of the origin. The two points are $\mu^{\dot \alpha} \bar \mu^{\alpha}$ and $\lambda^{\alpha} \bar \lambda^{\dot \alpha}$.

In the Lorentzian picture, the antiholomorphic diagonal $\mathbb{CP}^1_{[\mu \bar \lambda]=0}$ obtained from the $\mathbb{C}^*$ scaling reduction of $\{w=0\}$ is the space of points on the projectivised null cone of the origin in $\mathbb{M}$, the right playground for celestial holography. Geometrically, the locus $w=0$ is when $\mu^{\dot \alpha}$ is parallel to $\bar \lambda^{\dot \alpha}$, and therefore instead of an ordered pair of points $\mu^{\dot \alpha} \bar \mu^{\alpha}$ and $\lambda^{\alpha} \bar \lambda^{\dot \alpha}$ we have a single point $\mu^{\dot \alpha} \bar \mu^{\alpha} \sim \lambda^{\alpha} \bar \lambda^{\dot \alpha}$.

\paragraph{$\mathbb{PT}^\cO$ as a fibration over $\mathbb{MT}$ and $\mathbb{CP}^1_{[\mu \bar \lambda]=0}$:} On the patch of $\mathbb{PT}^{\cO}$ where $[\mu \bar \lambda] \neq 0$, we can coordinatise $\mu^{\dot \alpha}$ in the following way such that the only coordinates charged under the $\mathbb{C}^*$ are the $w$:
\begin{align}
    &\{w, \bar w, p, \bar p, \lambda^{\alpha}, \bar \lambda^{\dot \alpha}\}:=\left\{\frac{[\mu\bar \lambda]}{\la \lambda \hat \lambda \ra}, \frac{\la \bar \mu \lambda\ra}{\la \lambda \hat \lambda \ra}, \frac{[\mu \hat{\bar \lambda}]\la \lambda \hat \lambda \ra}{[\mu\bar \lambda]}, \frac{[\hat \mu \bar{\lambda}]\la \lambda \hat \lambda \ra}{\la \bar \mu \lambda\ra}, \lambda^{\alpha}, \bar \lambda^{\dot \alpha}\right\}, \quad p \in \mathcal{O}(2), w \neq 0 \nonumber
    \\
    &\mu^{\dot \alpha}=w \left(\hat{\bar \lambda}^{\dot \alpha}-p\frac{\bar \lambda^{\dot \alpha}}{\la \lambda \hat \lambda \ra}\right) \sim w \nu^{\dot \alpha}, \quad w\neq0, [\nu \bar \lambda] \neq 0\,,
\end{align}
where we have regrouped the inhomogenous coordinate $p, \bar p$ into the homogeneous coordinate $\nu^{\dot \alpha} \in \mathbb{CP}^1$ and promoted $w$ to $w \in \mathcal{O}(1_{\lambda},-1_{\nu})$ in each $\mathbb{CP}^1$ factor:
\begin{align}
     \{w, \bar w\}\times \{ \nu^{\dot \alpha},\bar \nu^{\alpha}, \lambda^{\alpha}, \bar \lambda^{\dot \alpha}\} \in \mathcal{O}(1,-1) \times \{\mathbb{CP}^1_{\lambda} \times \mathbb{CP}_{\nu}^1\}\setminus\{[\nu \bar \lambda] = 0\} \cong \mathcal{O}(1,-1) \times \mathbb{MT} \nonumber\,.
\end{align}
We see that the patch $\{[\mu \bar \lambda]\neq0\}$ trivially fibres over $\mathbb{MT}$, and the $\mathbb{C}^*$ action only acts on the fibre. In these coordinates, the basis of scaling eigenstates that will obey the square integrability constraint on $\mathbb{R}^4\setminus\{0\}$ is:
\begin{equation}
    Y_{\rho,m}(w, \bar w) = \left(\frac{w\bar w[\nu \hat \nu]}{\la \lambda \hat \lambda \ra}\right)^{\im \rho/2}\left(\frac{w \la \lambda \hat \lambda \ra}{\bar w[\nu \hat \nu]}\right)^{m/2} \in \mathcal{O}(m,-m), \quad m\in \mathbb{Z}, \rho \in \mathbb{R}\,.
\end{equation}
The decomposition into the basis of $\mathbb{C}^*$ scaling eigenstates therefore changes the weights of the fields:
\begin{equation}
    \kA(\mu,\lambda,\bar\mu,\bar\lambda)=\int \d \rho \sum_m Y_{\rho,m}(w, \bar w) \kA_{\rho,m}(\lambda, \nu,\bar\lambda,\bar\nu) \implies \kA_{\rho,m} \in \mathcal{O}(m,-m) \times \mathfrak{g}\,.
\end{equation}
On the patch where $[\mu \bar \lambda] = 0$, we coordinatise:
\begin{equation}
    \{q, \bar q, \lambda^{\alpha}, \bar \lambda^{\dot \alpha}\}:=\{[\mu \hat{\bar \lambda}], [\hat \mu \bar{\lambda}], \lambda^{\alpha}, \bar \lambda^{\dot \alpha}\}, \quad q \in \mathcal{O}(2)\,.
\end{equation}
This is a topologically trivial $\mathcal{O}(2)$ bundle over $\mathbb{CP}^1_{\lambda}$. The basis of $\mathbb{C}^*$ scaling eigenstates that obey the square integrability constraint on $\mathbb{R}^4\setminus\{0\}$ is:
\begin{equation}
    V_{\rho,m}(q, \bar q) = \left(\frac{q \bar q}{\la \lambda \hat \lambda\ra^2}\right)^{\im \rho/2}\left(\frac{q \la \lambda \hat \lambda\ra^2}{\bar q}\right)^{m/2} \in \mathcal{O}(2m), \quad m \in \mathbb{Z}, \rho \in \mathbb{R}\,.
\end{equation}
The mode decomposition into scaling eigenstates on the $\{[\mu \bar \lambda] = 0\}$ locus shifts the weights of the fields:
\begin{equation}
    \kA_{[\mu \bar \lambda] = 0}(\mu,\lambda,\bar\mu,\bar\lambda)=\int \d \rho \sum_m V_{\rho,m}(q, \bar q) \kA_{\rho,m}(\lambda, \bar\lambda) \implies \kA_{\rho,m} \in \mathcal{O}(-2m) \times \mathfrak{g}\,.
\end{equation}
In each of these cases, we have claimed a basis of scaling eigenstates that is required to obey a square integrability constraint on $\mathbb{R}^4\setminus\{0\}$. They are easiest understood in the language of local coordinates in the following section, in which we will see that they correspond to a $(1+\im \mathbb{R})$ Mellin basis in the radial coordinate (equivalent to Fourier transform in the log of the radial coordinate) and a Fourier series in a periodic angular coordinate on the $\mathbb{R}^4\setminus\{0\}$.
\paragraph{Large gauge transformations}
\label{appendix lgt}
What is a large gauge transformation? A large gauge transformation is required to be uncharged under the $\mathbb{R}^+$ ($\iff$ independent of $|x|=r$) and therefore the most generic thing (in the patch $\{[\mu \bar \lambda]\neq0\}$ for an Abelian positive helicity gluon) we can write is $a_f=\bar \partial f(\frac{w}{\bar w},p, \bar p, \lambda^{\alpha}, \bar \lambda^{\dot \alpha})$:
\begin{align}
    &\partial_{\alpha\dal} \Lambda = A_{\alpha\dal}(x)= -\frac{\partial}{\partial x^{\alpha\dal}} \left.\int\D\lambda\wedge\bar\delta(\la\lambda\,\iota\ra)\frac{\la\iota\,\hat \iota\ra}{\la\lambda\,\hat \iota\ra}f(1,p, \bar p, \lambda^{\alpha}, \bar \lambda^{\dot \alpha})\right\vert_{\mu=x\lambda}
    \\
    &\Lambda(x,\iota^{\alpha})=\tilde f\left(\frac{\la \iota| x| \hat{\bar \iota}]}{\la \iota |x |\bar \iota]},\frac{[\bar\iota |x| \iota\ra}{[\bar\iota |x|\iota\ra}  ,\iota^{\alpha}, \bar\iota^{\dal}\right)\,.
\end{align}
Note however that on real Minkowski space, $w/\bar w = 1$ and therefore the $w/\bar w$ dependence drops out of the integral formula. Therefore a large gauge parameter $\Lambda(x)$ on real Minkowski space corresponds to a function $f(p,\bar p ,\lambda, \bar \lambda)$ of $p, \lambda^{\alpha}$ and their complex conjugates. Note that $\bar \partial \tilde f$ will always contain a term with a distribution (delta function or derivative of a delta function) localising to a real codimension 2 locus of $\mathbb{PT}^{\cO}$, and for smooth $f$ the locus will be a formal neighborhood of $\mathbb{CP}^1_{[\mu \bar \lambda]=0}$. This is essentially because of the coordinate choice, in which we have picked the component of $\mu$ that is sampled by $[\mu \bar \lambda]$ as special and given ourselves a coordinate singularity at $[\mu \bar \lambda]=0$. We have picked it to be special because these are the coordinates suited for AdS$_3$ slices in the Milne region/dS$_3$ slices in the Rindler region and we are interpreting the CFT$_2$ as living on their common boundary, the projectivised null cone of the origin.

In the main text we consider gauge parameters with a pole of order 1 in $[\mu \bar\lambda]$, and then the subsequent localisation to the $\lambda$ sphere. We could have equally considered a pole of order 1 in $\la \bar \mu \lambda \ra$ localising to the $\mu$ sphere or higher order poles in each of these, which will give us information about formal neighborhoods off the S$^2$ and presumably, more information about bulk observables. This will appear in subsequent work.

\subsection{Details on mode decomposition}
We adopt the Euclidean perspective to do the $\mathbb{C}^*$ reduction, which divides $\mathbb{PT}^{\cO}\cong\mathbb{CP}^1\times\mathbb{R}^4\setminus\{0\}\cong \mathbb{CP}^1\times S^3\times\mathbb{R}_+$, where $S^3$ can be recognized as a Hopf fibration over $\mathbb{CP}^1$. As a preliminary, note that in the (scale-invariant) spacetime theory, we can rescale the spacetime fields and choose to work with the measure:
\begin{equation}
    \int \frac{\d^4 x}{x^4} = \int \frac{\d^4 x}{r^4}\,.
\end{equation}
By making the measure scale invariant, when we do a mode decomposition in the fields in the integrand there will be no constant offset in the sum of the mode numbers, which is clearer for presentation purposes. On twistor space, this amounts to absorbing factors of $[\mu \hat \mu]/\la\lambda \hat \lambda \ra$ into the $0,1$ forms that point in the $\mu$ direction:
\begin{equation}
    \text{conventional basis }\frac{\d\hat\mu^{\dot \alpha}}{\la \lambda \hat \lambda \ra} \rightarrow \frac{[\hat \mu \d \hat \mu]}{[\mu \hat \mu]^2}, \frac{[\mu \d \hat \mu]}{[\mu \hat \mu]}\,.
\end{equation}
Giving one of them a net $-2$ charge under real rescaling of $\mu$. The danger with working with such a basis is that $\frac{[\hat \mu \d \hat \mu]}{[\mu \hat \mu]^2}$ is not $\bar \partial$-closed, and care must be taken when decomposing covariant expressions into components with this basis choice. The measure on twistor space is now uncharged under the scalings:
\begin{equation}
    \frac{\la\lambda\hat \lambda \ra^2}{[\mu \hat \mu]^2}\D^3 Z\wedge \frac{\D^3 \bar Z}{\la \lambda \hat \lambda \ra^4} = \frac{1}{r^4} \d^4 x \wedge\frac{\D\lambda\wedge\D\hat{\lambda}}{\la\lambda\hat{\lambda}\ra^2}, \quad \text{as desired.}
\end{equation}
We work in the patch where $\mu^{\dot 1}\neq 0, \lambda ^0 \neq 0$. We begin by coordinatising $\mathbb{R}^4\setminus\{0\}$ with
\begin{equation}
    \frac{\mu^{\Dot{0}}}{\lambda_0} = \frac{rv\e^{\im\psi}}{\sqrt{1+|v|^2}},\quad
    \frac{\mu^{\Dot{1}}}{\lambda_0} = \frac{r\e^{\im\psi}}{\sqrt{1+|v|^2}}\,,
\end{equation}
and their complex conjugates $\bar\mu^{\Dot{0}}$, $\bar\mu^{\Dot{1}}$. Here $v=\mu^{\Dot{0}}/\mu^{\Dot{1}}$ denotes local coordinates on the Hopf fibration base $\mathbb{CP}^1$, $r\in\mathbb{R}_+$ denotes the radius of the $S^3$ and $\psi\in\mathbb{R}$ is the coordinate on the Hopf fibre $S^1$. We aim to dimensionally reduce on the $S^1$ and $\mathbb{R}_+$ factors. Firstly the holomorphic top form on $\mathbb{PT}^{\cO}$ can be written with this set of coordinates as:
\begin{equation}
    \D^3Z = \D\lambda\wedge\d\mu^{\Dot{0}}\wedge\d\mu^{\Dot{1}}= \frac{1}{2}\,\D\lambda\wedge\frac{\d v\,\e^{2\im\psi}}{1+|v|^2}\left(2r\d r+2\im r^2\d\psi-\frac{vr^2\d\bar v}{1+|v|^2}\right)\lambda_0\lambda_0\,.
\end{equation}
Similarly, the antiholomorphic top form can be worked out, collecting the top $(3,3)$ form in the new set of coordinates gives us:
\begin{equation}
    \frac{\la\lambda\hat \lambda \ra^2}{[\mu \hat \mu]^2}\D^3 Z\wedge \frac{\D^3 \bar Z}{\la \lambda \hat \lambda \ra^4}= \frac{\d v\wedge\d\bar v}{(1+|v|^2)^2}\wedge\frac{\D\lambda\wedge\D\hat{\lambda}}{\la\lambda\hat{\lambda}\ra^2}\wedge\d\psi\wedge\frac{\d r}{r}\,(\lambda_0\bar \lambda_0)^2\,.
\end{equation}
Note that although the coordinates $r^2=[\mu\hat{\mu}]$ and $\psi$ are not holomorphic on $\mathbb{PT}^{\cO}$, the remaining coordinates $\lambda$ and $v$ are on $\mathbb{MT}$. We also need to decompose the fields in \eqref{Split_action_PT} in $r$ and $\psi$. For example, for the redefined positive helicity gluon $\kA'$:
\begin{equation}
    \kA' = \sum_{m\in\mathbb{Z}}\int_{\mathbb{R}}\d \rho\,r^{\im \rho}\e^{\im m\psi}\,\kA_{m,\rho}(v,\lambda)\,.
\end{equation}
where we have used $\e^{\im m\psi}$ as a basis charged under the $S^1$ with charge $m\in\mathbb{Z}$. The Mellin transform in $r$ picks the charge under $\mathbb{R}_+$ to lie in the principle series $1+\im \rho$ (usual energy $\omega$ is the scaling of $r$). The basis choice for $r$ is forced on us by the condition of square integrability on the $\mathbb{R}^4\setminus\{0\}$ factor of $\mathbb{PT}^{\cO}$. This means that the scaling eigenstates we must use must have imaginary eigenvalues, and therefore we are taking the Fourier decomposition in $\log(r)$. This manifestly is equivalent to the Mellin decomposition in $r$ with Mellin modes $1+\im \mathbb{R}$. Note that instead of integer mode numbers $m$, our set up also allow $m\in\mathbb{R}$. This amounts to sections of nontrivial $\mathbb{C}^*$ bundles $A_{m,\rho}$ having non-trivial monodromy around the removed $\mu^{\dal}=0$ locus. In these Hopf coordinates, this corresponds to allowing the fields to wind around $r=0$ and pick up arbitrary phase. Although to reconstruct single valued 4d fields, we are required to take $m\in\mathbb{Z}$. 

Integration over $r, \psi$ for products of these basis elements restricts the mode numbers to sum to $0$. For example, we take the kinetic term in the $\mathbb{PT}^{\cO}$ action, collecting all the possible combinations of distributing the form degrees:
\begin{multline}
    \int_{\mathbb{PT}^{\cO}}\D^3Z\wedge \tr\left(\kB\wedge\bar\partial\kA \right)
    = \,\left(\int_{\mathbb{C}^*}\d\psi\wedge\frac{\d r}{r}\,\sum_{m,q\in\mathbb{Z}}\int_{(\mathbb{R}_+)^2}\d \rho\d\xi\,r^{\im(\rho+\xi)}\,\e^{\im(m+q)\psi}\right)\times\\ 
    \int(\lambda_0\bar \lambda_0)^2\frac{\d v\wedge\d\bar v}{(1+|v|^2)}\,\wedge\frac{\D\lambda\wedge\D\hat{\lambda}}{\la\lambda\hat{\lambda}\ra^2}\,\times \\
    \sum_{m,q\in\mathbb{Z}}\int_{(\mathbb{R}_+)^2}\d\rho\,\d \xi\,\tr\biggl(\left((\kB_{\bar w})_{m,\rho}\bar\partial_{\bar v} (\kA_{\bar \lambda})_{q,\xi}+(\kB_{\bar w})_{m,\rho}\bar\partial_{\bar \lambda}(\kA_{\bar v})_{q,\xi}-(\kB_{\bar v})_{m,\rho}(i\xi+q)(\kA_{\bar \lambda})_{q,\xi}\frac{\D\hat{\lambda}}{\la\lambda\hat{\lambda}\ra}\right)\,
    \\
    -\left((\kB_{\bar \lambda})_{m,\rho}\bar\partial_{\bar v} (\kA_{\bar w})_{q,\xi}+(\kB_{\bar v})_{m,\rho}\bar\partial_{\bar \lambda}(\kA_{\bar w})_{q,\xi}-(\kB_{\bar \lambda})_{m,\rho}(-2i+i\xi+q)(\kA_{\bar v})_{q,\xi}\frac{\D\hat{\lambda}}{\la\lambda\hat{\lambda}\ra}\right)\biggr),
\end{multline}
in which we label the components of $\kA$ (and $\kB$ likewise):
\begin{equation}
    \kA := \kA_{\bar \lambda} \frac{\D \hat \lambda}{\la \lambda \hat \lambda \ra^2} + \kA_{\bar v} \frac{[\hat \mu \d \hat \mu]}{[\mu \hat \mu]^2} + \kA_{\bar w} \frac{[\mu \d \hat \mu]}{[\mu \hat \mu]}\,.
\end{equation}
The mass term results from the $\bar \partial_{\bar w}$ derivative acting on the basis element $r^{\im \rho}e^{\im m \psi}$. The constant offset of $-2 \im$ is because $\frac{[\hat \mu \d \hat \mu]}{[\mu \hat \mu]^2}$ is not $\bar \partial$-closed. It is clear that we can now promote the local variable $v$ back into a homogeneous coordinate on $\mathbb{CP}^1$, amounting to including contributions from all 4 local coordinate patches of $\mathbb{PT}^{\cO}$. Integrating over $\mathbb{C}^*$ leaves an action on $\mathbb{CP}^1\times\mathbb{CP}^1$, after we appropriately re-pair the $(0,1)$ basis forms with their $\kA_{q,\xi}$ and $\kB_{m,\rho}$ components ($\kA_{\bar w},\kB_{\bar w}$ have lost their $(0,1)$ forms and they are relabelled as $(0,0)$ forms $\phi$ and $\eta$ respectively). The integration over $\mathbb{C}^*$ demands $m+q=0$ from the $S^1$ integral. The $\mathbb{R}_+$ integral can be dealt with after a change of variables $y=\e^r$, which reads
\begin{equation}
    \int_{-\infty}^{\infty} \e^{\im(\rho+\xi)y} \d y = \delta(\rho+\xi)\,,
\end{equation}
which we recognize as a distributional Dirac delta function. We see that integrating over the basis of $\mathbb{C}^*$ modes gives us constraints on the mode numbers of the component fields:
\begin{equation}
    m+q=0,\quad \rho+\xi=0\,.
\end{equation}
with these, we write the scaling reduced kinetic term as an action on $\mathbb{CP}^1 \times \mathbb{CP}^1$, after pairing the modes with antiholomorphic form degrees and relabelling the scalars as $\eta$ for unpaired $\kA_{-m,-\rho}$ and $\phi$ for unpaired $\kB_{m,\rho}$, we have
\begin{multline}
     \sum_{\Delta\in\mathbb{Z}+\im\mathbb{R}}\, \int_{\mathbb{CP}^1 \times \mathbb{CP}^1} \D\lambda\wedge\D\mu \,\tr\biggl(\eta_{\Delta}\bar\partial a_{-\Delta}+b_{\Delta}\wedge\bar\partial\phi_{-\Delta}-\Delta b_{\Delta} \wedge a_{-\Delta}+
     \\2 \im (b_{\lambda})_{\Delta}\wedge (a_{\mu})_{-\Delta}\frac{\D \hat \mu \D \hat \lambda}{[\mu \hat \mu]^2\la \lambda \hat \lambda \ra^2}\biggr)\,,
\end{multline}
where we have abused notation and denoted the minitwistor coordinates as $\mathbb{CP}^1_{\mu}$ and $\mathbb{CP}^1_{\lambda}$. $\Delta = (m+\im \rho)\in\mathbb{Z}+\im\mathbb{R}$ is some unified label for $\rho\in\mathbb{R}$ and $m\in\mathbb{Z}$. The $a$s and $b$s have the following form decompositions:
\begin{equation}
    a=a_{\mu}\frac{\D\hat{\mu}}{[\mu\hat{\mu}]^2}+a_\lambda \frac{\D\hat{\lambda}}{\la\lambda\hat{\lambda}\ra^2},\quad   b=b_{\mu}\frac{\D\hat{\mu}}{[\mu\hat{\mu}]^2}+b_\lambda \frac{\D\hat{\lambda}}{\la\lambda\hat{\lambda}\ra^2}\,.
\end{equation}
The reduction of the split action \eqref{Split_action_PT} can be done entirely analogously, obtaining \eqref{MT_split_action}.

\newpage
\bibliographystyle{JHEP}
\bibliography{CCFT}

\end{document}